\def\HA{{H$\alpha$}}

\def\LUM{\:{\rm ergs\:s^{-1}}}

\def\VEL{\:{\rm km\:s^{-1}}}

\documentclass[graybox]{svmult}

\usepackage{mathptmx}       
\usepackage{helvet}         
\usepackage{courier}        
\usepackage{type1cm}        
\usepackage{makeidx}         
\usepackage{graphicx}        
\usepackage{multicol}        
\usepackage[bottom]{footmisc}
\usepackage{deluxetable}
\usepackage{comment}

\makeindex             

\begin{document}


\newcommand{\MSOL}{\mbox{$\:M_{\odot}$}}  

\newcommand{\EXPN}[2]{\mbox{$#1\times 10^{#2}$}}
\newcommand{\EXPU}[3]{\mbox{\rm $#1 \times 10^{#2} \rm\:#3$}}  
\newcommand{\POW}[2]{\mbox{$\rm10^{#1}\rm\:#2$}}
\newcommand{\SING}[2]{#1$\thinspace \lambda $#2}
\newcommand{\MULT}[2]{#1$\thinspace \lambda \lambda $#2}
\newcommand{\CHINU}{\mbox{$\chi_{\nu}^2$}}
\newcommand{\vsini}{\mbox{$v\:\sin{(i)}$}}

\newcommand{\fuse}{{\em FUSE}}
\newcommand{\hst}{{\em HST}}
\newcommand{\iue}{{\em IUE}}
\newcommand{\euve}{{\em EUVE}}
\newcommand{\einstein}{{\em Einstein}}
\newcommand{\rosat}{{\em ROSAT}}
\newcommand{\chandra}{{\em Chandra}}
\newcommand{\xmm}{{\em XMM-Newton}}
\newcommand{\swift}{{\em Swift}}
\newcommand{\asca}{{\em ASCA}}
\newcommand{\galex}{{\em GALEX}}
\newcommand{\cxo}{CXO\,1337}
\newcommand{\etal}{$et\: al.$}

\title*{Galactic and Extragalactic Samples of Supernova Remnants: How They Are Identified and What They Tell Us}
\titlerunning{SNR Samples} 
\author{Knox S. Long}
\institute{Knox S. Long \at  Space Telescope Science Institute, 3700 San Martin Drive, Baltimore, MD, 21218 USA, also Eureka Scientific, Inc., 2452 Delmar Street, Suite 100, Oakland, CA, 94602 USA,  \email{long@stsci.edu}
}
%
%
\maketitle

\abstract*{Supernova remnants (SNRs) arise from the interaction between the ejecta of a supernova (SN) explosion and the surrounding circumstellar and interstellar medium.  Some SNRs, mostly nearby SNRs, can be studied in great detail.  However, to understand SNRs as a whole, large samples of SNRs must be assembled and studied. Here, we describe the radio, IR, optical, and X-ray techniques which have been used to identify and characterize almost 300 Galactic SNRs and more than 1200 extragalactic SNRs. We then discuss which types of SNRs are being found and which are not. We  examine the degree to which the luminosity functions, surface-brightness distributions and multi-wavelength comparisons of the samples can be interpreted  to determine the class properties of SNRs and describe efforts to establish the type of SN explosion associated with a SNR.  We conclude that in order to better understand the class properties of SNRs, it is more important to study (and obtain additional data on) the SNRs in galaxies with extant samples at multiple wavelength bands than it is to obtain samples of SNRs in other galaxies}

\abstract{Supernova remnants (SNRs) arise from the interaction between the ejecta of a supernova (SN) explosion and the surrounding circumstellar and interstellar medium.  Some SNRs, mostly nearby SNRs, can be studied in great detail.  However, to understand SNRs as a whole, large samples of SNRs must be assembled and studied. Here, we describe the radio, optical, and X-ray techniques which have been used to identify and characterize almost 300 Galactic SNRs and more than 1200 extragalactic SNRs. We then discuss which types of SNRs are being found and which are not. We  examine the degree to which the luminosity functions, surface-brightness distributions and multi-wavelength comparisons of the samples can be interpreted  to determine the class properties of SNRs and describe efforts to establish the type of SN explosion associated with a SNR.  We conclude that in order to better understand the class properties of SNRs, it is more important to study (and obtain additional data on)  the SNRs in galaxies with extant samples at multiple wavelength bands than it is to obtain samples of SNRs in other galaxies.}

\section{Introduction \label{Introduction}}

Supernova remnants (SNRs) are the visible manifestation of the interaction between material ejected in a supernova explosion (SN) and the surrounding circumstellar and interstellar medium. SNRs radiate across the entire magnetic spectrum from radio wavelengths to $\gamma$-rays. They provide the working surface where the elements produced in stars and supernovae (SNe)  and the kinetic energy of SN explosions mix with and stir the interstellar medium (ISM).  Shocks in SNR are responsible for the cosmic rays.  

SNRs are heterogenous.  The observational appearance of a SNR depends in a complex manner upon local factors such  the nature of the SN explosion, the presence or absence of an active pulsar, the time since the explosion, the mass loss history of the progenitor, the presence or absence of earlier SN, and the density and complexity of the surrounding medium.  Their appearance also depends on external factors such as the amount of absorption along the line of sight and the distance to the object.   

Samples of SNRs  provide an instantaneous picture of where stars are exploding in galaxies.  It is only a partial picture though, because many SNe explode in young massive star clusters, where other SNe have gone off recently.  Superbubbles, the emission nebulae created by the collective interaction of stellar winds and multiple SN from young star clusters on the ISM \cite{chu90}, are excluded from this discussion, and usually, though not always, have different observational characteristics.  For the purpose of this review, we define a SNR as the remnant of a single SN explosion. 

Although two SNRs - the Crab Nebula and Kepler's SNR - were identified earlier \cite{minkowski64}, the study to SNRs really began with the advent of radio astronomy, as it became clear that a significant number of bright sources in the plane of the Galaxy were indeed SNRs.  Today,  there are about 300 identified Galactic SNRs,  most within 90 degrees of the Galactic Center, and thus affected by interstellar absorption  \cite{green14}.  Most were first identified through their radio properties. The first extragalactic SNRs were identified in the Magellanic Clouds in the 1960's and 1970's \cite{mathewson63,mathewson73} through a combination of radio and optical techniques.  Since then  it has become possible to assemble large samples of samples of SNRs in galaxies out to a distance of about 10 Mpc.  Today there are about 59 SNRs and SNR candidates identified in the Large Magellanic Cloud (LMC) at distance 50 kpc \cite{maggi15},  217  in M33 at 812 kpc \cite{long10,lee_m33}, nearly 300 in M83 at 4.6 Mpc \cite{blair12,blair15}, and 93 in M101 at 6.7 Mpc \cite{matonick97}.  The total number of SNRs and credible SNR candidates in nearby galaxies exceeds 1200, four times the Galactic sample (see, e.g. the compilation of {Vu{\v c}eti{\'c}} \etal\ \cite{vucetic15} and Table \ref{table_extragalactic}).   With the exception of SNRs in the Magellanic Clouds, nearly all of the extragalactic SNRs have first  been identified optically.

The goals of research on SNRs are to understand what factors cause SNRs, individually and collectively, to appear as they do, and to separate the environmental factors from the astrophysics, such as the nature of the SN explosion and the effects of the explosion on the ISM as a whole.  Both the Galactic and the extragalactic samples are important in this regard.  

Because the SNRs in the Galactic and Magellanic Clouds samples are nearby and bright,  they provide the most direct confrontations of observations and theory.  
For example, in Cas A, where spectra of the light echoes from the explosion show the SN to have been of type IIb \cite{krause08}, Doppler imaging has allowed 3d reconstruction of the the ejecta at IR, optical and X-ray wavelengths \cite{fesen06, delaney10} and the spatial distribution of radioactive Ti from the explosion has been mapped \cite{grefenstette14}.  And in SN1006, where spatially resolved X-ray images were used to show that emission from the bright radio rims was synchrotron dominated and hence that SN shocks are capable of accelerating electrons to TeV energies \cite{koyama95}, high spatial resolution X-ray images obtained with \chandra\ are being used to limit the magnetic field amplification in the shock precursor \cite{winkler14}.    And, with a few exceptions, only in Galactic SNR is it possible to to identify pulsars and pulsar wind nebulae within a SNR \cite{gaensler06}, as seen, for example, in G292+1.8 \cite{park07}.

Extragalactic samples are also important:  First,  all of the SNRs observed in an external galaxy are effectively at the same distance and thus it is straightforward to translate observed fluxes and angular sizes to the physically more relevant quantities luminosity and diameter.  Second,  the effects of line of sight absorption on the appearance of a SNR are generally less severe and less variable than in the Galaxy, because one can choose to study external galaxies that are relatively face-on.  Third, it is easier, at least in principle, to account for observational selection effects in extragalactic samples because one can often conduct studies of SNRs in external galaxies with a single instrument at one time.   

The SNRs in the Magellanic Clouds merit special mention in terms of their utility; they are all at about the same distance along lines of sight with relatively little interstellar absorption so that it is fairly straightforward to examine them as a class, and close enough so that detailed multi-wavelength studies can be carried out of individual objects.

The purpose of this article is to describe how the SNRs in the Galaxy and external galaxies were and are continuing to be found, and to discuss the degree to which these samples are actually helping to address the goals of research on SNRs.  We will conclude that we have accumulated much useful information about SNRs as a class of objects, but that simple interpretations of the data, especially those that use diameter as a proxy for effective age, are naive.   Multifrequency studies, involving X-ray, optical, IR, and radio observations, of galaxies where SNR samples already exist are the best hope for gaining a more complete picture of SNRs as a class of objects.

\begin{deluxetable}{rrrr}
\tablecaption{SNRs and SNR Candidates in Nearby Galaxies}
\tablehead{\colhead{Galaxy} & 
 \colhead{Distance} & 
 \colhead{SNRs} & 
 \colhead{References} 
\\
\colhead{~} & 
 \colhead{(Mpc)} & 
 \colhead{~} & 
 \colhead{~} 
}
\tabletypesize{\scriptsize}
\tablewidth{0pt}\startdata
LMC &  0.05 &  53 &  1; \\ 
SMC &  0.06 &  25 &  2; \\ 
M31 &  0.79 &  156 &  3; \\ 
M33 &  0.84 &  217 &  4;5; \\ 
NGC300 &  2.00 &  22 &  6; \\ 
NGC4214 &  2.92 &  92 &  7; \\ 
NGC2403 &  3.22 &  150 &  8;9; \\ 
M82 &  3.53 &  50 &  10; \\ 
M81 &  3.63 &  41 &  8; \\ 
NGC3077 &  3.82 &  24 &  11; \\ 
NGC7793 &  3.91 &  27 &  12; \\ 
NGC4449 &  4.21 &  71 &  11; \\ 
M83 &  4.61 &  296 &  13;14; \\ 
NGC4395 &  4.61 &  47 &  11; \\ 
NGC5204 &  4.65 &  36 &  11; \\ 
NGC5585 &  5.70 &  5 &  15; \\ 
NGC6946 &  5.90 &  26 &  8; \\ 
M101 &  6.70 &  93 &  8;16; \\ 
M74 &  6.30 &  9 &  17; \\ 
NGC2903 &  8.90 &  5 &  18; \\ 
\tablenotetext{a}{ References: (1) Maggi \etal \cite{maggi15}; (2) Haberl \etal \cite{haberl12}; (3) Lee \& Lee \cite{lee_m31}; (4) Long \etal \cite{long10}; (5) Lee \& Lee \cite{lee_m33}; (6) Millar \etal\ \cite{millar11}; (7) Leonadaki \etal \cite{leonidaki13}; (8) Matonick \& Fesen \cite{matonick97}; (9) Leonidaki \etal\ \cite{leonidaki13}; (10) Huang \etal\cite{huang94}; (11) Leonidaki \etal \cite{leonidaki13}; (12) Blair \& Long \cite{blair97}; (13) Blair \etal \cite{blair12}; (14) Blair \etal  \cite{blair14}; (15) Matonick \& Fesen \cite{matonick97}; (16) Franchetti \etal \cite{franchetti12}; (17) Sonba{\c s} \etal \cite{sonbas10}; (18) Sonba{\c s} \etal  \cite{sonbas09};}
\enddata 
\label{table_extragalactic}
\end{deluxetable}

\begin{figure}[b]
\includegraphics[scale=.6]{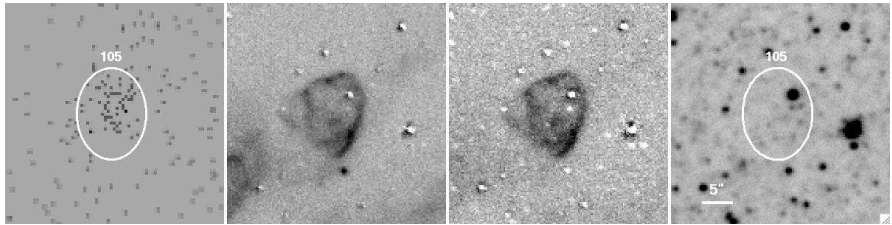}
\caption{An example of a SNR in M33 as described by Long \etal\ \cite {long10}. From left to right, the panels show the field of the SNR as observed in X-rays with \chandra, and in  \HA, [S II] and the V-band continuum as observed in ground-based images from  the Local Group Galaxy Survey of Massey \etal\  \cite{massey06}.
Stars have been subtracted from the emission line images.  Notice that the \HA\ region seen in the lower left corner of the \HA\ fades compared the the SNR in the [S II] image.  Had this object not been known as a SNR as a result of the optical observations, it would have been discovered as an X-ray SNR due to its soft spectrum and spatial extent in the X-ray image.
}  
\label{fig_m33_example}       
\end{figure}

\section{Techniques for finding SNRs and SNR candidates}

Most of the SNRs in the Galaxy were initially identified as extended radio sources with non-thermal radio spectra.  However, most extragalactic SNRs, and SNR candidates,  an example of which is shown in Fig.\ \ref{fig_m33_example},  have been identified optically using narrow band imaging.  Progress has been rapid, due to the development of CCD detectors, which coupled with the angular resolution of optical telescopes, allowed one to isolate SNR candidates from H II regions.  X-ray and radio discovery of SNRs in external galaxies has largely, though not exclusively, been limited to the Magellanic Clouds, where limitations associated with angular resolution and sensitivity are less severe.  Some progress in detecting SNRs in X-rays has been made with the launch of \chandra\ and XMM-Newton and in the radio with  the increasing sensitivity of the Jansky Very Large Array (JVLA) and the Multi Element Radio Linked Interferometer Network (MERLIN). The most useful studies of SNRs, especially in galaxies beyond the Magellanic Clouds, will be those that involve observations in at least these three wavebands, so it is important to pursue each of them vigorously.

\subsection{Optical Identification of SNRs \index{Optical Identification of SNRs} \label{optical}}

\begin{figure}[b]
\includegraphics[scale=.4]{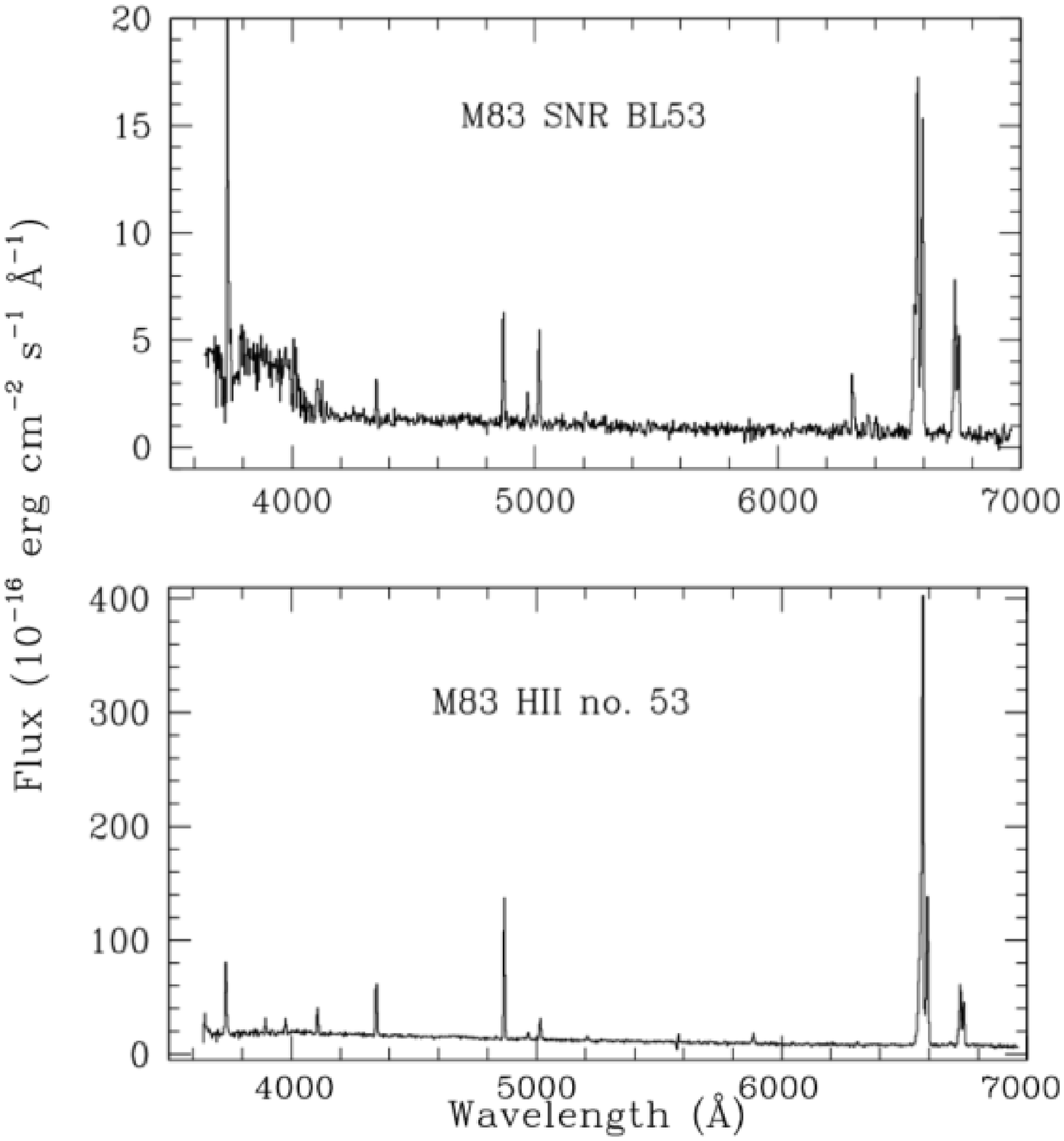}
 \caption{The spectra of a typical SNR candidate  and a bright H II region in M83 as observed by Blair \& Long \cite{blair04}.  The SNR shows much more \MULT{[S II]}{6717,6731} compared to \HA\ than the H II region, as well as emission from \MULT{[O I]}{6300,6363} and \MULT{[N II]}{6549,6583}.  The quality of the spectra is also fairly typical of those observers try to obtain to confirm line ratios from imaging observations.  }
\label{fig_opt}       
\end{figure}

Optically,  SNRs are extended sources, which must be distinguished from the other type of emission nebulae -- H II regions -- that exist in galaxies.  In SNRs,  optical emission  normally arises from shocks, most commonly from radiative shocks driven into relatively dense clouds in the ISM by the primary shock wave.  These secondary shocks, with typical velocities $v$ of 200 $\VEL$,  heat the post-shock gas to a temperature of order 500,000 $(v/200 \VEL)^2$ K, ionizing it to a degree which depends on the shock velocity.  However, at these temperatures, the plasma radiates very efficiently.  As a result, gas cools behind the shock, increasing further in density, recombining to the neutral state on a timescale that is short compared to the cloud crossing time. As a consequence, models predict \cite{dopita77,raymond79,allen08} and observations show optical spectra containing forbidden lines from a wide range of ionization states , including in the optical, \MULT{[O III]}{4959,5007}, \MULT{[O I]}{6300,6363},  \MULT{[N II]}{6549,6583}, and \MULT{[S II]}{6717,6731}.  

Unlike SNRs, the optical emission in H II regions arises from gas photoionized by UV photons from hot stars. In H II regions, most of the optical emission is produced by recombination and emerges in the the Balmer lines.  Most of the material in  H II regions is too highly ionized to produce  forbidden lines of O I, S II, and N II.  Furthermore, at least in bright H II regions,  there is a sharp boundary between fully ionized gas inside the so-called Str{\"o}mgren sphere, and unionized gas in the region outside the sphere, so there is relatively little gas at intermediate ionization states.  As a result, as shown in Fig.\  \ref{fig_opt}, the spectra of SNRs and H II regions differ.

First suggested as a technique by Mathewson \& Clark \cite{mathewson73}, essentially all SNRs that have been identified optically in external galaxies have been identified  as emission nebulae with elevated [S II]:\HA\ ratios compared to H II regions. In bright H II regions, the [S II]:\HA\  is typically about 0.1, whereas in SNRs the ratio is typically 0.4 or greater.  Searches are conducted using interference filter imaging, with filters centered on \HA\ (often also including a contribution from [NII]), [S II] and a continuum band.   One inspects these images for emission nebulae that show elevated [S II] compared to \HA, designating as candidates extended objects with [S II]:\HA $>0.4$. (Occasionally, slightly lower or higher values have been used.)  An example of a SNR discovered in this way is shown in Fig.\ \ref{fig_m33_example}. The SNR is recognized by the fact that it is relatively much brighter in the [S II] image than the H II region in the lower left hand corner.  Often, follow-up spectroscopy is carried out, which not only confirms the   [S II]:\HA\  ratios, but also allows searches for additional SNR indicators,  usually   [O I] emission, and in rare cases, velocity broadening of the lines. 

The technique works best for isolated SNRs, where one can measure the ratio of [SII]:\HA\ emission without the diluting effect of an adjacent/underlying H II region,  and for high surface brightness nebulae. Lower surface brightness H II regions tend to have higher [S II]:\HA\ ratios.  In particularly, Blair \& Long \cite{blair97} found that in NGC7793, the [S II]:\HA\ ratio often exceeds 0.5 in nebulae with surface brightnesses of less than \POW{-15}{erg~cm^{-2} s^{-1}arcsec^{-2}}.  The Str{\"o}mgren sphere is simply not as well defined in low density H II regions and the ionization levels drop more slowly with distance from the ionizing stars than with higher density, so that there is an extended region where ions, such as S II, are prevalent.

Partly to address these problems, many observers eliminate from consideration nebulae with high [S II]:\HA\ ratios with obvious evidence of a concentration of blue stars.  The advantage of this strategy is that it makes it more likely that an object identified as a SNR, actually is a SNR. The disadvantage is that SN do explode in regions with blue stars, and one's candidate list is less complete.  

Observers also have to decide whether 
to include or exclude nebulae which satisfy  the [S II]:\HA\  test, but which are larger than expected from a single SN with a typical explosion energy.  Many of these  objects are, as argued recently by Franchetti \etal\ \cite{franchetti12},  superbubbles or collections of SNRs.   Consequently, some observers have excluded objects larger than (typically) 100 pc from SNR candidate lists \cite{lee_m33,lee_m31}. Others have retained them \cite{matonick97,long10}, feeling any particular diameter arbitrary and arguing that over time the reality or not of any particular candidate with be determined by future observations.

Although the optical emission from most SNRs arises from radiative shocks in gas with near interstellar abundances,  several other types of optical emission are observed less commonly in SNRs: 

(a) \index{Balmer-dominated SNRs}  A small number of SNRs exist, notably SN1006 and Tycho's SNR, which radiate only in the Balmer lines \cite{chevalier78,raymond10,winkler14}.  In these SNRs, the optical emission arises from a so-called non-radiative shock, in which a fast shock, typically $>$1,000 $\VEL$, encounters a partially neutral ISM.   In these situations, the cooling time behind the shock is long compared to the age of the SNR, and the only optical radiation arises as the plasma is ionizing.  
The surface brightness of the optical emission from these Balmer-dominated SNRs is low compared to those that that emit via  radiative shocks, and the spectra are not easy to distinguish from H II regions.  Consequently, the only extragalactic SNRs to have been identified of this type have been in the LMC \cite{tuohy82}, objects which were first detected as X-ray sources \cite{long81}.  A few remnants of this type continue to be discovered in the Galaxy, including recently G70.0-21.5 \cite{fesen15}.  In principle, such objects could be discovered in other galaxies if observed with sufficient spectral resolution to detect large velocity broadening; in practice, it is more likely that a Balmer-dominated SNR will be identified first in another wavelength range.

(b) SNRs also exist in which line emission arises from interactions with the ejecta from core-collapse SNe, such as is the case for the Galactic SNRs,  Cas A \cite{kirshner77} and G292+1.5 \cite{goss79}.  Optical emission from the ejecta of such SNe is characterized by very strong emission from forbidden lines of O II and O III, which are very efficient coolants for a plasma with abundances expected in the ejecta of core-collapse objects \cite{dopita84}.  A number of searches for SNRs of this type have been carried out.  A few objects have been found, e. g. E0102-72.9 in the SMC \cite{finkelstein06} and the remnants of some very young SNe, such as SN1957D in M83 \cite{long89} and the very bright SNR in NGC4449 \cite{kirshner80}.  The numbers are, however, very small, and all of these objects were first discovered by other means.  Optical searches for these SNRs are difficult because they are expected to be small diameter objects, and easy to confuse with planetary nebulae and certain stars with strong emission lines..

(c) Finally, there are some SNRs, often referred to a pulsar wind nebulae, \index{Pulsar wind nebulae} where optical line emission arises from circumstellar material/ejecta photoionized  by synchrotron radiation due ultimately to the active pulsar.  Unlike the photoionization produced by thermal emission from hot stars, the hard power law synchrotron spectrum is capable of leaving the plasma in a large variety of ionization states, which results in emission line spectra that look significantly different from a normal H II region.  In principle, such SNRs could be discovered through measurements of the [S II]:\HA\ ratio, or could be buried in existing catalogs of extragalactic planetary nebulae. To date, none has been recognized  beyond the Magellanic Clouds, with the possible exception of SN1957D in M83 \cite{long12}.

\begin{figure}[b]
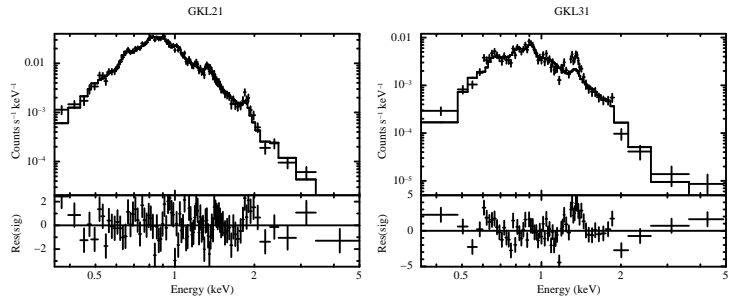

\includegraphics[scale=.2,angle=-90,origin=c]{fig3a.eps}
\includegraphics[scale=.2,angle=-90,origin=c]{fig3b.eps}

 \caption{X-ray spectra of the two brightest X-ray SNRs in M33 as observed with \chandra\ by Long \etal\ \cite{long10}.  The spectra show clear evidence of the line emission expected from a shocked thermal plasma.}
\label{fig_m33_spectra}       
\end{figure}

\subsection{Radio identification of SNRs \index{Radio identification of SNRs} \label{radio}}

At radio wavelengths, SNRs in the Galaxy are extended, non-thermal radio sources. Shell-like SNRs in particular typically have radio spectral indices $\nu^{-\alpha}$ of about 0.5 though with considerable dispersion (see, e. g. Fig. 6 of Dubner \& Giaconi \cite{dubner15}).  H II regions, which are also extended sources at radio wavelengths, are the main source of confusion.  These are thermal radio sources, radiating primarily by free-free emission, which has a spectral index of 0.1. This means that shell-like SNRs can in principle be separated from H II regions if the spectral index can be measured.   The pulsar-dominated SNRs, like the Crab Nebulae, have flatter spectral indices from 0.0 to 0.3 \cite{kargaltsev15} and are harder to identify on this basis, but these constitute a relatively small portion of the total sample of the Galactic sample, and would presumably be a similarly small portion of any complete extragalactic sample as well.

Not surprisingly the first extragalactic radio SNRs to be identified/detected are located in the Magellanic Clouds \cite{mathewson63,mathewson73}. Indeed, with the availability of the Australia Telescope Compact Array (ATCA), all known SNRs in the Large and Small Magellanic Clouds have been detected at radio wavelengths  \cite{maggi15,filipovic05}.  

Identification at radio wavelengths  of SNRs in more distant galaxies has been hampered by a number of factors.  
First, until very recently radio observations did not generally have the combination of sensitivity and angular resolution necessary to detect and measure the spectral indices of potential SNR candidates in more distant galaxies. Secondly, SNRs are often found in regions with other diffuse emission and this can dilute the spectral index of putative SNRs, especially in the absence of multi-frequency maps with the same spatial resolution.  Finally, as surveys have grown more sensitive contamination from background sources has become an issue, particularly in Local Group galaxies, which have substantial angular diameters.  Consequently, most of the radio-detected SNR candidates are sources which were identified as SNR candidates optically and then detected as radio sources (see, e. g. Gordon \etal\ \cite{gordon99} for the case of M33).  

There have been some searches particularly in galaxies outside the Local Group where observers have identified as SNRs non-thermal radio sources with associated \HA\ emission (see, e.g. \cite{lacey01,chomiuk_search}).  This mitigates the background source problem, and is positive in the sense that it does not depend an optical SNR identification, but it also introduces spatial biases into the sample that are difficult to quantify.  Objects such as RXJ1713-39 \cite{pfeffermann96}  and RX J0852.0-4622 (also
known as Vela Junior) \cite{aschenbach98}, which have no associated \HA\ emission, and  the historical SNRs, SN1006 and Tycho, which are very faint in \HA\ would almost certainly be missed.  

Lacey \& Duric \cite{lacey01} used this approach to identify 35 radio point sources in NGC6946 as SNRs, almost none of which were in Matonick \& Fesen's \cite{matonick97} list of 27 optical SNR candidates. The radio sample in NGC6946 has radio fluxes corresponding to 0.1-2 times that of Cas A, and is systematically brighter than the radio-detected optical sample in M33. The radio candidates are more closely associated with bright H II regions and with the spiral arms than the optically identified SNRs, which Lacey \& Duric  suggest is at least partially due to observational biases associated with identification of optical SNR candidates.    They suspect SNRs in  the radio sample in NGC6946 are evolving in denser interstellar environments than SNRs in the optical sample.\index{NGC6946}

\subsection{X-ray identification of SNRs \index{X-ray identification of SNRs} \label{xray}}

\begin{figure}[b]
\includegraphics[scale=.38]{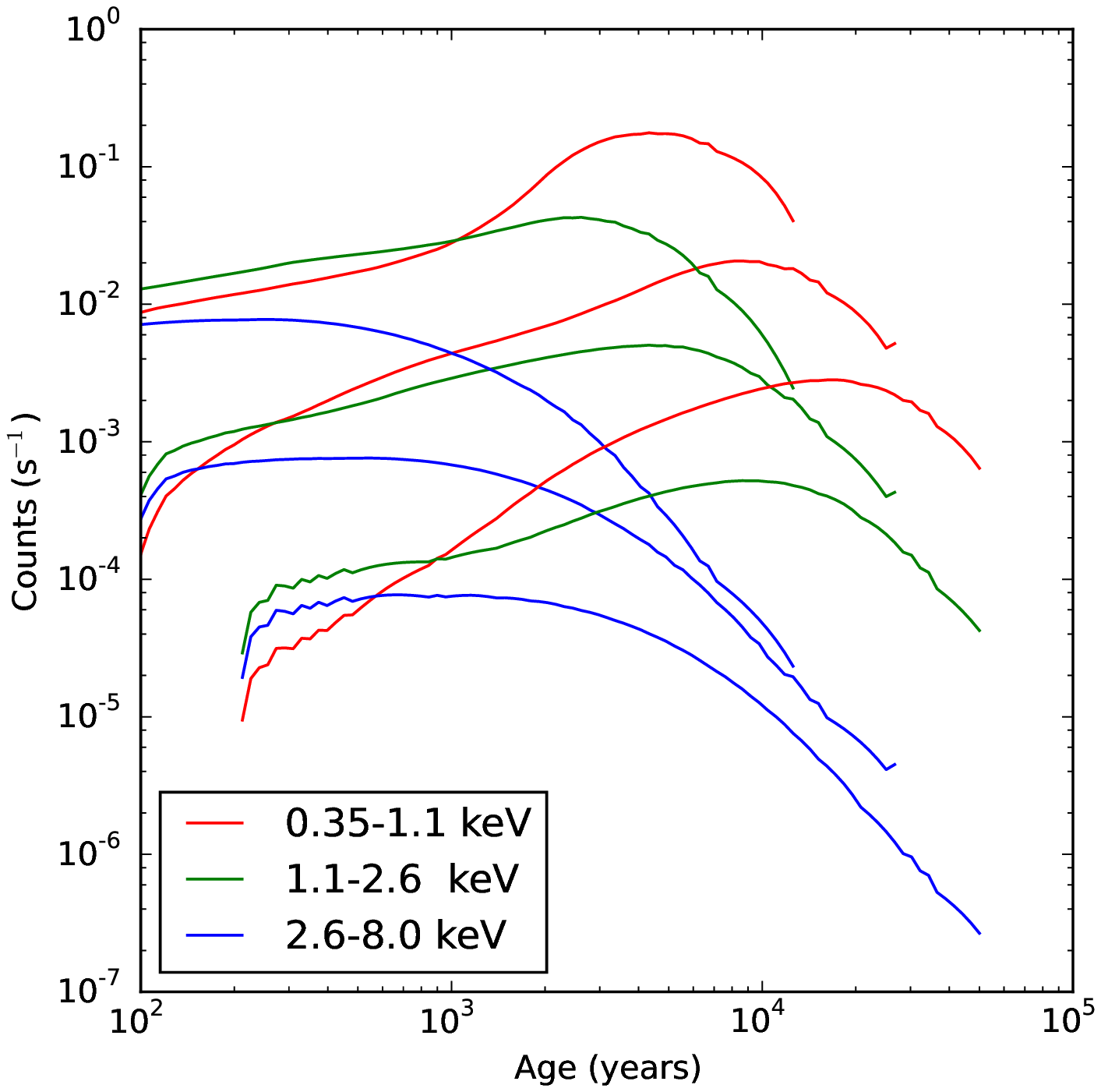}
\includegraphics[scale=.38]{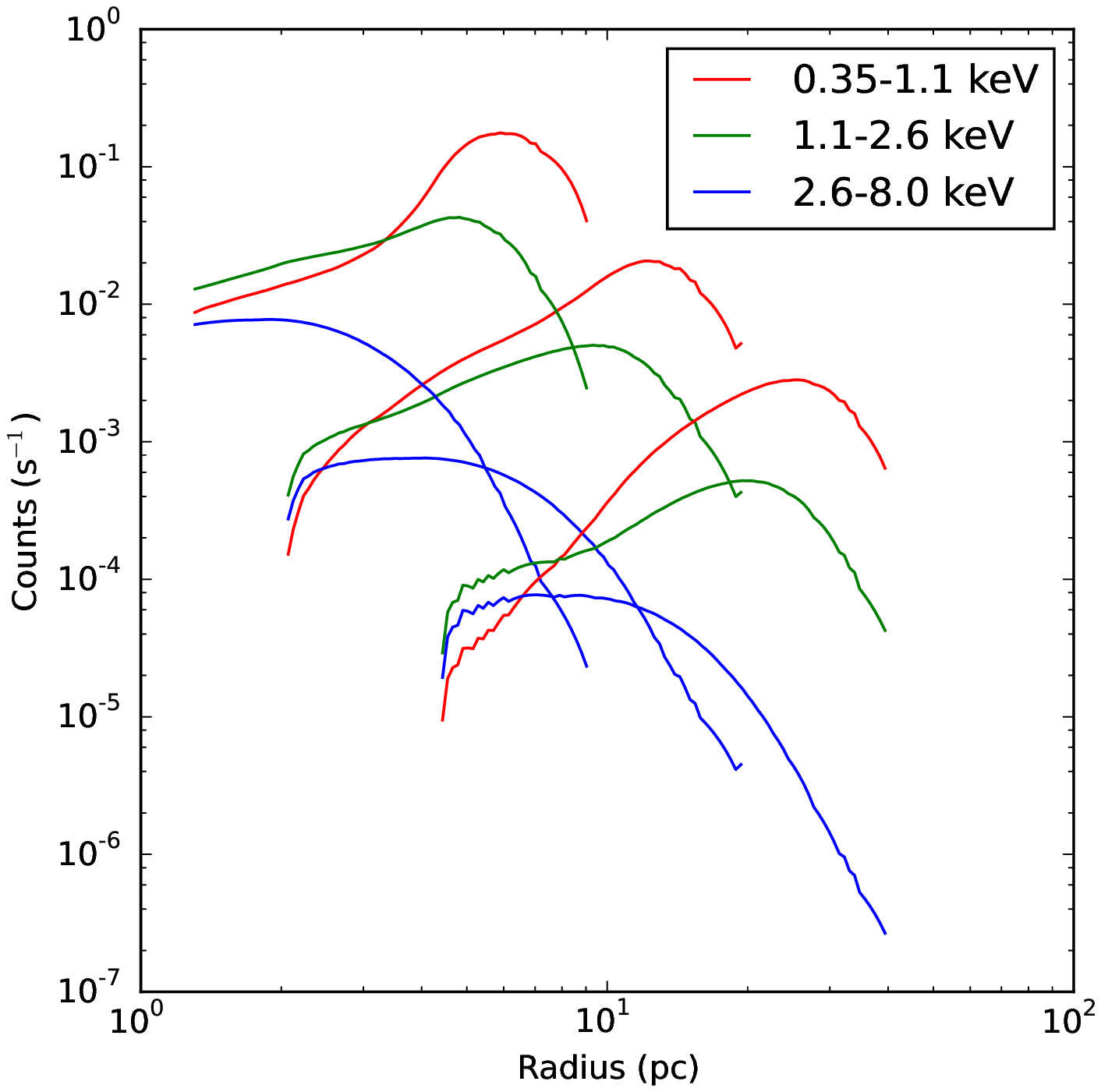}

 \caption{Predicted \chandra\ count rates in 3 bands (0.35-1.1 keV, 1.1-2.6 keV, and 2.6-8 keV) for SNRs at a distance of 1 Mpc assuming they are in the Sedov phase and assuming a line of sight absorption of \EXPU{5}{20}{cm^{-2}}. All of the curves terminate at a kT of about 0.09 keV (because {\sc XSPEC} models do not exist at lower kT); this however, is well before the beginning of the radiative phase.}
\label{fig_xray_sensitivity}       
\end{figure}

SNRs, as indicated in the leftmost panel of Fig.\ \ref{fig_m33_example}, are also extended sources at X-ray wavelengths.  Most have soft, line dominated X-ray spectra, arising from hot \EXPN{1}{6} to \EXPU{5}{7}{K} gas produced by the reverse shock interaction with SN ejecta or the primary shock interaction with the ISM. Even if the shocks speeds are high enough to produce a plasma hotter than this, the spectra looks as if it has a temperature in this range due to ionization equilibration effects.  \chandra\ spectra obtained by Long \etal\ \cite{long10} of the two brightest SNRs in M33 are shown in Fig.\   \ref{fig_m33_spectra}.   A small number of SNRs, those powered by pulsars, such as the Crab and 3C58 in the Galaxy,  and a few young synchrotron-dominated, SNRs, such as RXJ1713-39 and Vela Jr, have power law spectra.  

To give an indication of what one expects to see from an X-ray SNR in a nearby galaxy, we show, in Fig.\ \ref{fig_xray_sensitivity}, estimated  \chandra\ count rates for SNRs in the Sedov phase at a distance of 1 Mpc as function of age and size, as calculated with the program {\sc XSPEC}, a routine used widely in the astrophysics community to fit X-ray spectra \cite{xspec}. The three sets of curves are for SNRs expanding into ISM with densities of 10, 1, and 0.1 cm$^{-3}$.  These rates are indicative of a number of important ``facts'' about the expected detectability of SNRs in X-rays.  SNRs brighten through much of their Sedov phase and are easiest to detect at ages of 10,000-20,000 years.  In the early Sedov phase, SNRs are relatively faint because they have not swept up enough material; in the late Sedov phase; they fade because the post-shock plasma temperature has dropped.  SNRs expanding into a dense ISM are brighter, but evolve more rapidly.  The spectra are soft, at least at ages greater than 1,000 years.  

H II regions also contain thermal plasma, but they typically are much less luminous  ($<10^{34} \LUM$) than SNRs.  Giant H II regions do have X-ray luminosities of up to \POW{37}{\LUM}, but are easy to isolate based on their \HA\ luminosities and stellar content.   Wind blown bubbles  are occasionally not distinguishable from SNRs, especially if one allows large objects ($>$100 pc diameter) in the sample.

Although most Galactic SNRs were first identified as radio sources, a number were first detected or suggested as  SNRs as a result of their detection as extended X-ray sources from  the ROSAT all-sky survey\cite{schaudel02}.  These include the aforementioned RXJ1713-39  and Vela Jr, as well as more typical thermal plasma dominated objects, such as G38.7+1.4 \cite{huang14}, G296.7-0.9 \cite{robbins12}, G299.2-2.9 \cite{busser96},  and G308-1.4 \cite{hui12}.   It is quite likely that additional Galactic SNRs will be identified as part of the eROSITA all-sky survey  \cite{erosita14}, which will be about 30 times more sensitive than ROSAT.

In other galaxies, especially those beyond the Local Group, SNRs are relatively faint and difficult or impossible distinguish from point sources on the basis of spatial extent.  Fortunately, most  other  galactic X-ray sources (neutron star and black hole binaries)  and most background sources (AGN and galaxy clusters) have relatively featureless hard spectra that with 
spectral resolution and counting statistics are easy to distinguish from the thermal plasma-dominated spectra of most SNRs, simply on the basis of hardness ratios.  However, this still leaves a group of compact sources, the so-called supersoft sources thought to be white dwarf binaries  that have  luminosities as high as \POW{38}{\LUM}.  These objects have very soft hardness ratios (corresponding to effective temperatures of \POW{5}{} to \POW{6}{K}) \cite{kahabka97}, which makes them hard to separate from SNRs (given limited source counts).  Stiele \etal\ \cite{stiele11} identify 30 sources in their survey of M31 with XMM as supersoft sources, which is comparable to the number of objects they suggest are SNRs.  Many of these supersoft sources are variable, but the fact that this source population exists means that it is dangerous to assume that all soft X-ray sources in a galaxy are SNRs.  Consequently, most observers require something other than a hardness ratio to declare an X-ray source as a SNR candidate,  usually association with an optical or radio source that has the properties of a SNR.

There have been a few X-ray only searches for SNRs, where observers have tried to establish a set of X-ray candidates without complementary information at other wavelengths.  Leonadaki \etal\ \cite{leonidaki10} identified 37 objects in archival \chandra\ data as SNR candidates in six nearby galaxies (NGC2403, NGC3077, NGC4214, NGC4449, NGC4395, and NGC5204) based  on their X-ray hardness ratios or colors.  Of these, only 7 had been previously suggested as SNR candidates.  This is a useful exercise since one is not biased by the characteristics expected for SNRs at other wavelengths, though it remains to be seen how many of these objects actually turn out so be SNRs as more sensitive observations are carried out.

\subsection{IR Identification of SNRs \index{IR identification of SNRs}}

Historically, very few SNRs have been identified via observations in the IR.  However, the [Fe II] $\lambda\lambda$1.27,1.64 $\mu$m is, like [S II], a tracer of radiative shocks, and can be used in conjunction with, for example, Pa$\beta$ to separate H II regions from SNRs (see, e.g. Mouri \etal\ \cite{mouri00} for a discussion of the shock models and Oliva \etal\ \cite{oliva89,oliva90} for early IR spectroscopy of Galactic SNRs ).  

Greenhouse \etal\ \cite{greenhouse97} used Fabry-Perot imaging  of the [Fe II] $\lambda$1.64 $\mu$m  line to identify 6 sources in M83 which they argued were an older population of SNRs than those identified in the radio.  Subsequently, Morel \etal\ \cite{morel02} imaged 42 [S II]-identified SNRs in M33, detecting about 10, with [Fe II]   $\lambda$1.64 $\mu$m luminosities of \EXPU{0.2-27}{35}{\LUM}.  The advantage of NIR imaging is that line of sight absorption is less of a problem in the IR; the disadvantages are that the night sky is much more of a problem in the IR and until recently, advances in detector technology were delayed compared to CCDs.  

\begin{figure}[b]
\includegraphics[scale=.38]{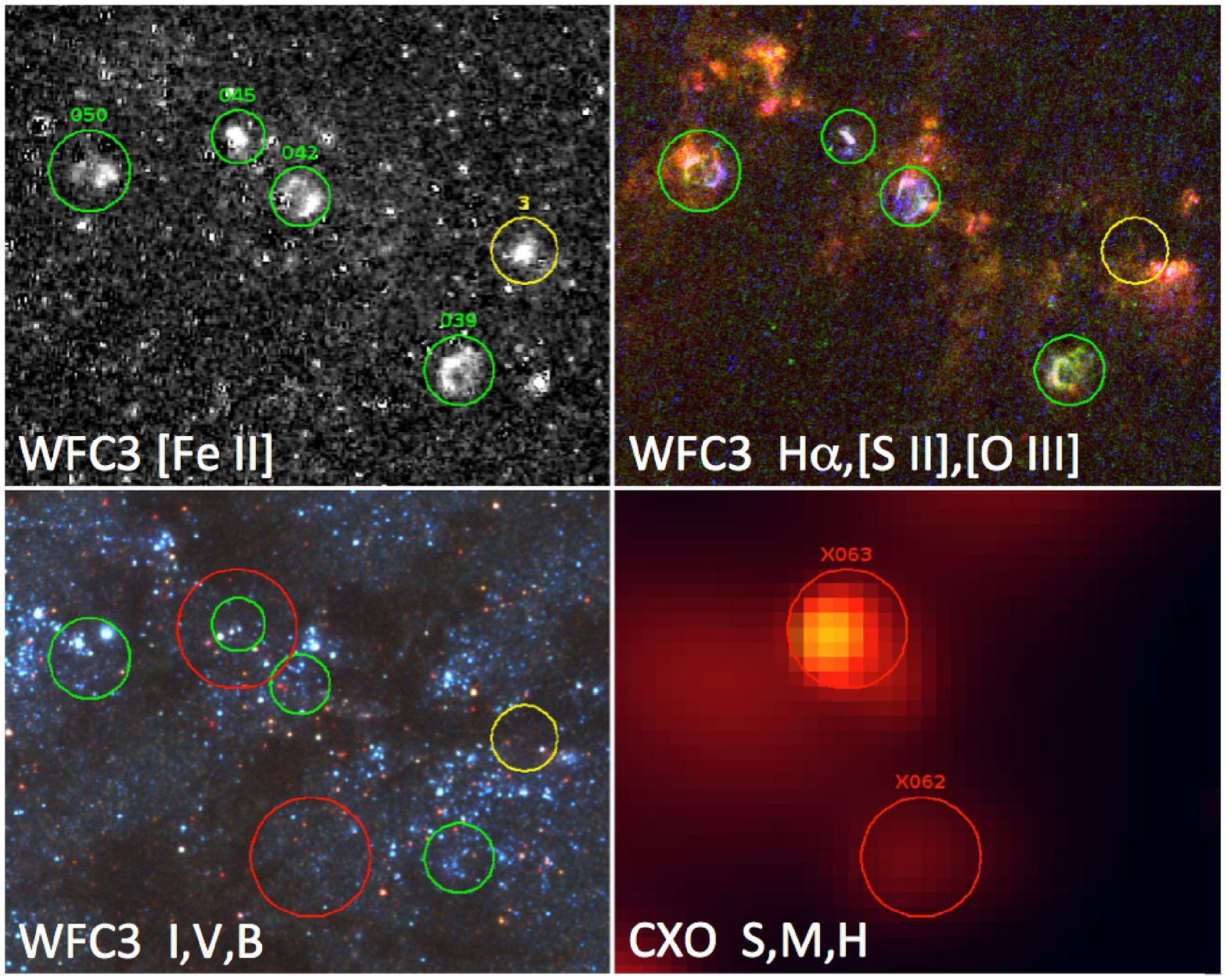}

\caption{A 16" by 20" portion of M83 as observed with HST and \chandra \cite{blair15}.  The upper left panel shows the [Fe II] image, the upper right panel shows a composite \HA, [SII], [O III] image; the lower left panel shows a composite of U, B, V images; and the lower light panel shows the \chandra\ image.  Objects identified as SNRs from [S II] imaging by Blair \etal\ \cite{blair12} are shown in green; X-ray sources from the catalog of Long \etal\ \cite{long14} are shown red, one of which the X-ray counterpart to an optical SNR.  All of the SNRs have [Fe II] counterparts.  The field also contains one other object, identified in yellow, that is bright in [Fe II] and is most likely a SNR behind a dust lane. }
\label{fig_fe2}
\end{figure}

However, the picture is changing with improvements in IR detectors.  Blair \etal\ \cite{blair15}, as part of of Hubble Space Telescope (HST) imaging study of M83, has found a number of emission nebulae in M83 coincident with X-ray sources that are apparent in [Fe II] $\lambda$1.64 $\mu$m, but not in [S II], that  are likely to be SNRs.  An example is shown in Fig.\ \ref{fig_fe2}.  
Some impetus for such studies should arise both from the more systematic studies of [Fe II] emission in Galactic SNRs that are currently underway \cite{lee14}, and from assertions that the [Fe II] imaging of galaxies can be used an an estimator of the SN rates \cite{rosenberg12}.

\section{The Samples Today\label{Indiviidual}}

\subsection{The Galaxy \index{Galactic SNR sample}}


According to Green \cite{green14}, there are now 294 identified SNRs in the Galaxy. Nearly all have been detected at radio wavelengths, while 40\% have been detected in X-rays and 30\% have been detected in the optical (the low fraction being a result of the effects of absorption in the plane of the Galaxy).  Of the SNRs, 79\% are shell-like (though in many cases the actual shell-structure is quite complex), and 5\% are center-filled (dominated by emission from the ``wind'' of  a central pulsar like the Crab).  The remaining SNRs have a composite morphology, with evidence for emission from both the central pulsar and a shell.  Usually, SNRs with radio shells detected at X-ray wavelengths also show X-ray shells, but there is a group comprising some 20\% of the total that have center-filled X-ray morphologies.  

Mid-IR emission, which arises both from shock-heated dust grains and from IR lines in hot gas, from SNRs was first detected in the all sky survey conducted with the Infrared Astronomical Satellite (IRAS) at 12, 25, 60, and 160 $\mu$.   Arendt \cite{arendt89} and later Saken \etal\ \cite{saken92} claimed detections of about 30\% of the SNRs known at the time.  They found that the morphologies of SNRs in the mid-IR were similar to that observed at X-ray and radio wavelengths and established that the IR luminosities of SNRs were in some cases comparable to their X-ray luminosities.  With the Spitzer Space Telescope, new surveys of SNRs were undertaken with considerably higher precision. In particular, {Pinheiro Gon{\c c}alves} \etal\ \cite{pinheiro11} found 39 counterparts to the 121 SNRs contained in the region surveyed with Multiband Imaging Photometer for Spitzer (MIPS) as part of the MIPSGAL Survey (at 24, 60, and 160 $\mu$); they argued that the detection rate was primarily limited by confusion in the plane of the Galaxy, that X-ray bright SNRs were found preferentially and that the IR luminosities of the detected SNRs were comparable to X-ray luminosities.   At shorter Infrared Array Camera (IRAC) wavelengths (3.6, 4.5, 5.8, and 8.0 $\mu$), where emission from SNRs can arise from shock-heated dust, atomic fine-structure lines, molecular lines and occasionally synchrotron emission, Reach \etal\ \cite{reach06} reported 18 detections in the GLIMPSE survey region containing 95 SNRs.  Many of the other SNRs in the survey region  could have significant infrared emission but are located along lines of sight with large amount of emission due to H II regions and atomic and molecular clouds.

At least 30 of the Galactic SNRs have been detected in $\gamma$-rays (1-100 GeV) with Fermi LAT (see Acero \etal\ \cite{acero15} for a good summary of the current list of detections, and an overall interpretation). \index{Gamma-ray detection of SNRs}  The detected SNRs, which have typical $\nu L_{\nu}$ luminosities of \POW{35}{\LUM},  fall into two subclasses, ``young SNRs" in the free expansion or Sedov phases, and SNRs interacting with molecular clouds.  $\gamma$-rays  can be produced from relativistic electrons either through inverse Compton emission or by bremsstrahlung radiation, or alternatively from relativistic protons (and other hadrons) which created pions which then decay to $\gamma$ rays.  Both processes are thought to play a role.  To date, a correlation between the radio and $\gamma$-ray fluxes has not been demonstrated \cite{acero15}.  

The sample of SNRs in the Galaxy is not really complete, at least in the sense that all SNRs of certain intrinsic properties in the Galaxy have been discovered.  Green \cite{green14_distribution} estimates that the radio sample is approximately complete to a surface brightness limit of \POW{-20}{Watts~m^{-2}Hz^{-1}sr^{-1}}. There are 68 SNRs brighter than this limit, but Green notes that a selection bias still exists. Specifically, it is  hard to recognize small diameter, distant SNRs which will be in the Galactic plane along a line of sight near the Galactic Center where source confusion is likely.  The situation is clearly worse at optical and X-ray wavelengths, where absorption is more of a problem.   

The number of Galactic SNRs should continue to grow with better all sky surveys at X-ray wavelengths, such as eROSITA \cite{erosita14}, and emission line surveys, including the Isaac Newton Telescope Photometric \HA\ Survey  IPHAS \cite{IPHAS,sabin13}, its southern hemisphere VLT counterpart VPHAS \cite{VPHAS},  and the UKIRT wide field imaging survey of Fe+ (UWIFE) \cite{UWIFE}.  The follow-up from detections of $\gamma$-ray sources is also likely to continue to pay dividends in this regard.

\subsection{Magellanic Clouds \index{Magellanic Clouds}}

Because the Large and Small Magellanic Clouds are nearby (50 and 60 kpc, respectively) and because they lie along lines of sight with very low Galactic and internal absorption, more is known about the SNRs in the Magellanic Cloud as a group than in any other galaxy.  
Not only is it fairly straightforward to study the SNRs, it is also possible to study the environments around them as result of the large amount of ancillary data that has been accumulated on the Magellanic Clouds for a variety of other purposes. 

The first SNRs in the LMC were detected as non-thermal radio sources by Mathewson \& and Healey \cite{mathewson64} and subsequently confirmed as SNRs on the basis of strong [SII]:\HA\ ratios by Westerlund \& Mathewson \cite{westerlund66}.    The numbers grew during the 1970s, primarily as a result of work by Mathewson \& Clarke \cite{mathewson73}, as radio and optical instrumentation became more sensitive and as systematic searches for SNRs were carried out.  Long, Helfand \& Grabelsky \cite{long81} used \einstein\ to carry out the first X-ray imaging survey of the LMC; of the 97 X-ray sources detected, they found 26 SNRs, including a number which had not been known previously.    

According to Maggi \etal\ \cite{maggi15}, there are currently 59 confirmed SNRs in the LMC. \index{Large Magellanic Cloud} Nearly all have been detected at X-ray, optical, and radio wavelengths.  Most of the SNRs have optical spectra.  Russell \& Dopita \cite{russell90} (see, also \cite{payne08})  have used the spectra to measure ISM  abundances in the LMC.  Echelle spectra exist for a significant fraction of the SNRs, allowing one to study the expansion velocity of the optical filaments \cite{chu97}.  One of these SNRs B0540-69.3 \cite{mathewson80,brantseg14}  contains an 80 ms pulsar \cite{seward84} which produces $\gamma$-ray pulses 20x brighter than the Crab pulsar \cite{LMC_pulsar}.  One young SNR, N49, is coincident with a soft $\gamma$-ray repeater \cite{cline80,kulkarni03,guver12}.  Light echoes from the SN explosion have been seen from three \cite{rest05}.  At least one X-ray bright SNR, N132D, has portions of its optical spectrum dominated by emission from the shocked ejecta \cite{danziger76,vogt11}

About 60\% of the SNRs known in the LMC have been detected with Spitzer \cite{seok13}  in the NIR where emission can arise from molecular shocks, synchrotron radiation, ionic lines, or PAH emission and/or in the MIR (mainly) from shock-heated dust.  Seok \etal\ \cite{seok13} use these data to argue that LMC SNRs are fainter on average than Galactic SNRs in the IR, presumably due to lower dust to gas ratios in the LMC than the Galaxy, and that the SNRs of Type Ia SNe are significantly fainter than those arising from core-collapse SNe.  

This situation is similar for the Small Magellanic Cloud (SMC) \index{Small Magellanic Cloud} although the total number of SNRs (25) is smaller, as one would expect since the SMC is less massive the LMC. The first optical/radio SNR in the SMC was discovered by Mathewson \& Clarke \cite{mathewson72}); the first X-ray SNR. the second brightest X-ray source in the SMC, was found by Seward \& Mitchell \cite{seward81}, as part of the first X-ray imaging survey carried out with the \einstein\ Observatory.  
Nearly all of the SMC SNRs have been detected in X-rays \cite{haberl12} with XMM, and most have radio fluxes \cite{filipovic05,payne07} and optical spectra \cite{russell90,payne07}.  Many of the SNRs have been studied in detail. The remnant E0102-72.3, discovered with \einstein, is one of the very small number of SNRs showing  emission from ejecta at optical wavelengths \cite{blair00}.  One SNR, HFPC 334, appears to be a composite SNR, comprised of an active pulsar inside a shell-like radio source \cite{crawford14}.

\subsection{M33 \index{M33} \label{m33}}

The first three SNRs in M33 were identified by {D'Odorico}, {Benvenuti}, \& {Sabbadin} \cite{dodorico78} using interference filters and image tube photography, and the numbers grew significantly with the advent of CCDs \cite{long90,gordon98}.  

A detailed study of SNRs in M33 using a combination of deep \chandra\ exposures and optical data from the Local Group Galaxy Survey (LGGS) \cite{massey06}
survey was carried out by Long \etal\ \cite{long10}.  They found 137 optical SNR candidates (with [S II]:\HA $>$0.4) in M33, with diameters ranging from 8 to 179 pc. Of these,  82 were detected in X-rays with 0.35-2 keV luminosities in excess of \EXPU{2}{35}{\LUM}, and of these seven were bright enough for detailed spectral analysis.  Based on a spectral  analysis of all of the sources detected in M33, Long \etal\ estimated that they had identified all of the thermal plasma-dominated X-ray SNRs brighter than \EXPU{4}{35}{\LUM}, at least in the region covered by the \chandra\ survey.

Subsequently, Lee \& Lee \cite{lee_m33} reexamined the LGGS data and produced a larger sample of 199 optical SNR candidates.   Their sample is larger in part because they surveyed a larger region of M33 and pushed to fainter surface brightnesses, and somewhat different because they excluded objects with diameters greater than 100 pc.  They argued that objects with these characteristics were unlikely to be SNRs and should be excluded. Long \etal\ discussed some of these concerns, but felt that excluding such objects as candidates was premature.   However, the differences also reflect the subjective aspects of identifying optical SNR candidates, the majority of which are very faint and near the sky background limit in the LGGS and other ground based data.  If the two lists are combined, there are 217 optically-identified SNRs and SNR candidates in M33.  Of these, 86 of these, all from the list of Long \etal, have optical spectra.   

Most recently, Williams \etal\ \cite{williams15} have described the results of their analysis of a new deep set of XMM observations covering a larger region of M33 than was observed with \chandra; in addition to recovering most of the SNRs reported as X-ray sources by Long \etal, they detected 8 new X-ray SNRs, three of which are in the outskirts of the galaxy.

{D'Odorico}, Goss \& Doptia \cite{dodorico82} carried out the first successful radio search for SNRs in M33 using the Westerbork Synthesis Radio Telescope (WSRT) at 21 cm.  They reported  five certain and three probable detections of sources at the positions of the  12 optically-identified SNRs known at the time.  
Subsequently, Gordon \etal\ \cite{gordon99} used the a combination of Very Large Array (VLA) and WSRT observations obtained at 6 and 20 cm with an angular resolution of 7", or 30 pc, to construct a catalog of 186 sources in M33.  Of these sources,  they identified 53 sources as spatially coincident with one of the 98 optically-identified SNRs known at this later date.  The mean radio spectral index of the radio sources identified as SNRs was 0.5, and the summed radio luminosity of SNRs in M33 comprised 2-3\% of the total synchrotron emission in M33.  There were a number of other non-thermal sources detected above their surface brightness limit of 0.2 mJy along the line of sight to M33, but they noted that most of these were likely background sources.  

None of  SNRs identified in M33 is very young.  There are no objects, like N132D or E0102-72.3, in the Magellanic Clouds, whose optical emission is dominated by emission from shocked ejecta, or even any SNRs with broad optical emission lines.  There is an  X-ray source with a power-law spectrum coincident with a small-diameter radio source that Long \etal\ suggest may be a pulsar-wind nebula.

\subsection{M31 \index{M31} \label{m31}}

The first few SNRs in M31 were identified Kumar \cite{kumar76}, and subsequent image tube photography by D'Odorico \etal\ \cite{dodorico80} and by Blair \etal\ \cite{blair81}, respectively,  expanded the number of spectroscopically confirmed SNRs to 14.
Because of its very large size, M31 was actually less surveyed than several other nearby galaxies for many years.  Braun \& Walterbos \cite{braun93} found 52 SNR candidates in the first CCD-based search for SNRs in M31, but surveyed only a portion of the galaxy.  Magnier \etal\ \cite{magnier95} identified 179 candidates in seventeen fields  totaling a square degree of the galaxy, but their selection of candidates was based on morphology in \HA. They did not use the [SII]:\HA\ ratio as a criterion, and as a result, there was no direct evidence that the majority of  the nebulae selected by Magnier \etal\  actually contained shocks.  The situation has changed recently however, as Lee \& Lee \cite{lee_m31} have searched the LGGS survey images of M31 for SNRs, just as they had done for M33.  They  identified 156 emission nebulae with diameters less than 100 pc as SNRs or SNR candidates on the basis of [S II]:H{$\alpha$} $>$0.4 and circular morphology.  Most of the candidates are associated with the spiral arms of M31.

Although the first X-ray detection of a SNR in M31 was most likely made by Blair \etal\  \cite{blair81} using \einstein,
the first reliable characterization of the X-ray properties of SNRs in M31 has required the greater sensitivity of \chandra\ and especially XMM \cite{pietsch05, stiele11}.  According to Sasaki \etal\ \cite{sasaki12}, there are now 26 confirmed X-ray SNRs in M31, 21 of which had been thought to be SNRs based on earlier observations, and six which were discovered as a result of the XMM studies.  These SNRs are confirmed in the sense that they have both the X-ray and optical characteristics of SNRs. The X-ray luminosities of the SNRs range from \EXPU{2}{35}{\LUM} to \EXPU{8}{36}{\LUM} in the 0.3-2 keV band.  There are also 20 candidate SNR, objects that either have soft X-ray spectra, but ambiguous evidence from other wavelength bands as to whether the object is a SNR, or hard X-ray spectra but evidence for a radio source or a nebula with high [S~II]:\HA\ ratios at that position.  

The first radio search for SNRs in M31 was carried out Dickel \etal\ \cite{dickel82} who used the VLA at 20 cm and reported the radio detection of 7 SNRs identified earlier by {D'Odorico} \etal\ \cite{dodorico80}. 
Although some other efforts to characterize small numbers of SNRs in M31 have taken place since then \cite{braun93,sjouwerman01}, the SNR population of M31 is still not well-characterized at radio wavelenths. Galvin \& Filipovic \cite{galvin14} have published a catalog of 916 point sources in 20 cm radio images of M31 constructed  from archival VLA data and compared the positions of these point sources to SNR candidates suggested by others.  With a flux limit of about 2 mJy, 
they find 13 objects whose position matches those contained in the list of optical candidates produced by Lee \& Lee \cite{lee_m31}.  Of the 47 SNRs and SNR candidates reported by Sasaki \etal\ \cite{sasaki12} in X-rays with XMM, they find 11 overlaps.  

As is true of M33, no very young SNRs have been identified as yet, with the exception of the remnant of SN 1885, which Fesen \etal\ \cite{fesen15_m31} have imaged in absorption with HST against the stars in bulge of M31. 

\subsection{Supernova Remnants Beyond the Local Group}

The first 17  SNR candidates in six galaxies beyond the Local Group were identified by {D'Odorico} \etal\ \cite{dodorico80} using photogaphic plates.  The first large  CCD-based searches were carried out by Blair \& Long \cite{blair97} who identified 56 SNR candidates in the Sculptor Group Galaxies, NGC300 and NGC7793,  and by Matonick \& Fesen \cite{matonick97} who identified a total of about 400  SNR candidates in NGC5204, NGC5585, NGC6946, M81, and M101.   

As shown in Table \ref{table_extragalactic}, optical samples of varying depths now exist for more than 20 galaxies  within 10 Mpc \cite{vucetic15}.  These include NGC2403 with 150 candidates \cite{leonidaki13}, M83 with nearly 300 candidates \cite{blair12,blair14},  and M101 with 93 candidates \cite{matonick97}.  A necessary next step for improving the reliability of these samples is to obtain the spectra of as many of these candidates as possible. This is underway for many of these galaxies, including M81, where Lee \etal\ \cite{lee15_m81} have obtained spectra of 28 of 41 optically identified SNR candidates; they find that 26 of their 28 objects should be retained as candidates.

As noted earlier, there have been relatively few dedicated radio searches  for  SNRs outside of the Local Group. \index{Radio identification of SNRs} In the Sculptor group spiral NGC300, Payne \etal\ \cite{payne04} used data from the VLA and from ATCA to identify 18 non-thermal radio sources associated with \HA\ emission or an X-ray point source in XMM data as SNRs; five of these were in Blair \& Long's list of optical SNRs, but 13 were new.   Of the 18 sources, six were also detected with XMM.   In another Sculptor group spiral NGC7793, Pannuti \etal\ \cite{pannuti02} identified five radio SNR candidates.  Lacey  \etal\  \cite{lacey01}, as discussed earlier, identified 35 objects in the starburst galaxy NGC6946 as radio SNRs, six of which Pannuti \etal\  \cite{pannuti07} found to be X-ray sources in \chandra\ images.  

M82 \index{M82} is an exception in terms of the importance of radio observations.  At 3.2 Mpc, M82 is the closest example of a prototypical starburst galaxy,  that is a galaxy  undergoing a huge burst of star formation (due in the case of M82 to a near collision with M81). Such galaxies contain a large amount of dust, and this dust, heated by early type stars, radiates strongly in the FIR.  The current star formation rate (SFR) in M82 is about 10 $\MSOL ~yr^{-1}$, much greater than the Galaxy.  With this SFR,  M82 produces a large number of SNe, 1 every 10 to 20 years,  and hence a large number of very young SNRs.  The SNRs are mostly buried behind large amounts of dust and hence primarily accessible at radio wavelengths.   The first radio studies of  M82 were carried out by Kronberg \& Wilkinson \cite{kronberg85}, and the galaxy has been monitored since that time with the VLA and MERLIN.  Huang \etal\ \cite{huang94} observed the highly reddened  galaxy  with the VLA at a resolution of 0.2" and identified 50 sources near the center of the galaxy with diameters all less than that of Cas A, and with higher radio surface brightness as well.  They argued  the vast majority of these were SNRs expanding into the high pressure ISM of the central region of M82.  They found these sources obey a $\Sigma$-D relationship that extrapolates to that of the $\Sigma$-D of Galactic and Magellanic Cloud SNRs.  Repeated observations with Merlin and the VLA and with very long baseline interferometry have allowed one to measure the time evolution of the radio fluxes and, in many cases, the expansion velocities of the SNRs.  For example, Fenech \etal\ \cite{fenech08} used MERLIN to detect about 35 SNR in M82 ranging in diameter from 0.3 to 6.7 pc, with a mean of 2.9 pc. Most of the sources show shell-like morphologies.  They measured expansion velocities ranging from 2200 $\VEL$ to 10,500 $\VEL$ in 10 SNRs.  These velocities are significantly larger than predicted by Chevalier \& Franson \cite{chevalier01} who suggested that the radio SNRs in M82 were expanding into a dense \POW{3}{cm^{-3}} ISM and mostly in their radiative phase.  The distribution of diameters in this SNRs measured by Fenech \etal\ suggests that most of the SNRs are in the free expansion phase and that the SNRs are expanding into the region carved out by the winds of a progenitor red giant star.

X-ray identifications of SNR candidates in galaxies beyond the Local Group have mostly proceeded from attempts to identify X-ray sources with nebulae satisfying the optical criteria for SNRs or as radio SNRs. Deep observations with \chandra\ are required to see all but the most luminous SNRs in Galaxies beyond the Local Group. \index{X-ray identification of SNRs}  Matonick \&  Fesen \cite{matonick97} had identified 93 emission nebulae as optical SNR candidates in M101. Franchetti \etal\  \cite{franchetti12} re-examined the 55 objects in the sample that were contained in archival \HA\ images obtained with HST, ranging in size from 20 to 330 pc, including 16 with diameters greater than 100 pc.  They found that these that 21 of the 55 candidates had X-ray counterparts in very deep (1 Ms) \chandra\ observations \cite{kuntz10}.   And Long \etal\ \cite{long14} analyzed a series of \chandra\ observations of M83 totaling 729 ks.  They found 378 point sources within the D$_{25}$ contour of the galaxy, including 87 sources which appeared to be SNRs based on a combination of their X-ray properties and coincidence with either an optical SNR candidate or a radio source within the galaxy.\index{M83}   Smaller numbers of X-ray detected SNRs exist in other galaxies outside the Local Group.

Not surprisingly, most of the SNRs that have been discovered in galaxies beyond the Local Group appear, with the limited information available, to be older, larger diameter SNRs, since one typically looks for objects with some indication of spatial extent, and small diameter SNRs radiating primarily in the oxygen lines are hard to separate from planetary nebulae.  There have been nine SNe in NGC6946, and six in M83 in the last 100 years, which implies there are about 90 and 60 SNRs in these galaxies of age less than 1,000 years. But this also means there are many more SNRs with ages of up to 20,000 years, so young SNRs are going to be rare.   A few may have been found depending on one's decision about when to declare an object a SNR, as opposed to late time emission from a SN since some SNe have been observed essentially continuously from the time they exploded \cite{danny12}.   The bright optical, radio, and X-ray SNR in NGC4449 was probably due to a SN that was missed in the last 50-100 years \cite{danny08}.  

One pathway forward to identifying younger SNRs is through higher angular resolution observations at optical or radio wavelengths. There are now some searches that are being carried out with the WFC3 on HST, which has the requisite filters, that might address this problem.  For example, in M83, Blair \etal\ \cite{blair14} found 26 new small diameter SNRs (by carefully inspecting HST images near X-ray point sources)  and measured small ($\leq$ 0.5 arcsec or 10 pc) diameters for 37 others.  However, so far only one of these objects is known to have very broad optical emission lines \cite{blair15}; this particular object, which was also detected as an X-ray source and a radio source,  was most likely another example a SN in the last 100 years that was missed.  The rest of the small diameter objects look to be SNRs that are evolving in a denser ISM than the typical SNR in other galaxies.

\section{What the samples tell us about SNRs as a class}

The past 50 years have seen a lot of progress in terms of identifying samples of SNRs both in the Galaxy and nearby galaxies. However, we now need to ask what we can learn from these samples.  

\subsection{Luminosity Function \index{SNR luminosity function}}

\begin{figure}[b]
\includegraphics[scale=.38]{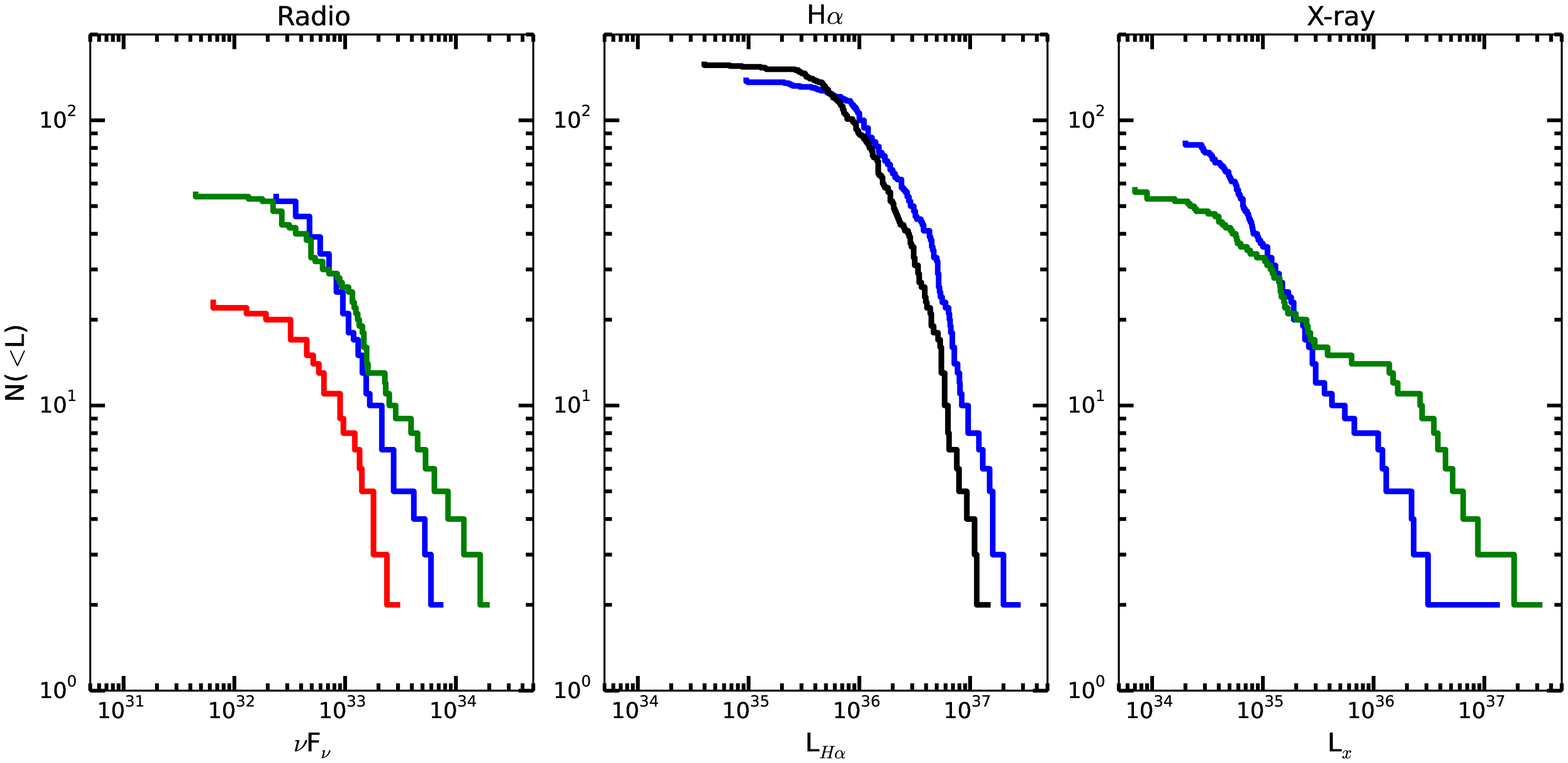}
\caption{The radio, \HA, and X-ray luminosity functions of SNRs in several nearby galaxies.  The various luminosity functions M31, M33, and the Large and Small Magellanic Clouds are shown in blue, black, green and red, respectively.  Data for M33 was taken from Gordon \etal\ \cite{gordon99} and Long \etal\ \cite{long10}; for M31 from Lee \etal\ \cite{lee_m31} and for the Large and Small Magellanic Clouds from Badenes \etal\ \cite{badenes10} and Maggi \etal\ \cite{maggi15}.  The radio luminosity shown is the specific luminosity at 20 cm.}
\label{fig_lumfunc}
\end{figure}

The luminosity function of SNRs\index{SNR luminosity function}, expressed as the number of SNRs with a luminosity less than a specific value,  at radio, optical (\HA) and X-ray wavelengths for several galaxies is shown in Fig.\ \ref{fig_lumfunc}.  In all cases, the shapes of the luminosity functions are affected by sample completeness at low luminosity.
To the extent that conditions are similar in these galaxies, one might  expect that the normalization of the luminosity functions would reflect the number of SNe in each galaxy, and, since most SNe arise from relatively young stellar populations, the overall SFR.  Leonidaki \etal\ \cite{leonidaki10}, for example, assert that  the number of X-ray SNRs brighter than \POW{36}{\LUM} in a galaxy is proportional to the SFR.  They also suggest that SNRs in irregular galaxies tend to be more luminous than in spiral galaxies, possibly due to the fact that the lower metal abundances observed in irregular galaxies results in more massive SN progenitors.  Maggi \etal\ \cite{maggi15} find the X-ray luminosity functions of both M31 and M33 can be fitted as a power law with a slope about 0.8, but that the SMC is significantly flatter (0.5).  They find that the luminosity function of the LMC is complex, with 13 SNRs brighter than \POW{36}{\LUM}.  Like Leonidaki \etal, they attribute this due to low metallicity, but suggest that the winds of more massive stars in the LMC create low density cavities, which result in luminous SNRs when the SN shock reaches the cavity walls at ages of a few thousand years \cite{dwarkadas05}. \index{M31} \index{M33} \index{Magellanic Clouds}

Similarly, Chomiuk \& Wilcots \cite{chomiuk09}, using a sample of radio SNR candidates in 18 galaxies, ranging from the Magellanic Clouds to galaxies like M51 and M82,  argue SNRs generally have a power distribution of luminosities with a scaling that is proportional to the SFR.   Specifically, they find overall
\begin{equation}
\frac{dN}{d L_{\nu} }= 92 ~ SFR~ L_{\nu} ^{-2.02}
\end{equation}
where $L_{\nu} $ is the specific luminosity at 1.4 GHz (20 cm) in mJy and  SFR is  in units $\MSOL~yr^{-1}$.  Unlike Thompson \etal\ \cite{thompson09}, they find no indication that the peak (or average) luminosity is related to the gas density of the galaxy.

At optical and X-ray wavelengths, luminosity functions are difficult to interpret in terms of physical models because the amount of emission is very dependent on the local ISM density. However, this may not be the case at radio wavelengths.  Following early theoretical work by Reynolds \& Chevalier \cite{reynolds81} and {Berezhko} \& {V{\"o}lk} \cite{berezhko04}, Chomiuk \& Wilson argue  that the slope of the radio luminosity function can be understood in terms of a model in which (a) most of the SNRs are in the Sedov phase, (b) the cosmic ray energy is a fixed fraction of the SN explosion energy throughout the Sedov phase, and (c) the magnetic field energy density behind the shock is amplified to $\sim 0.01 \rho_o v_s^2$.   

Whether this particular interpretation of the radio luminosity function is actually physically correct is difficult to determine; as the specific luminosities of SNRs with the same diameter vary substantially (see below), so a complete interpretation of the radio emission from SNRs has to account for these differences.  Nevertheless, luminosity functions  at all wavelengths are  clearly useful for estimating completeness of samples, and the total amount of radiation arising from SNRs at the various wavelengths.

\subsection{The Diameter Distribution of SNRs \index{SNR diameter distribution}}

\begin{figure}[b]
\includegraphics[scale=.38]{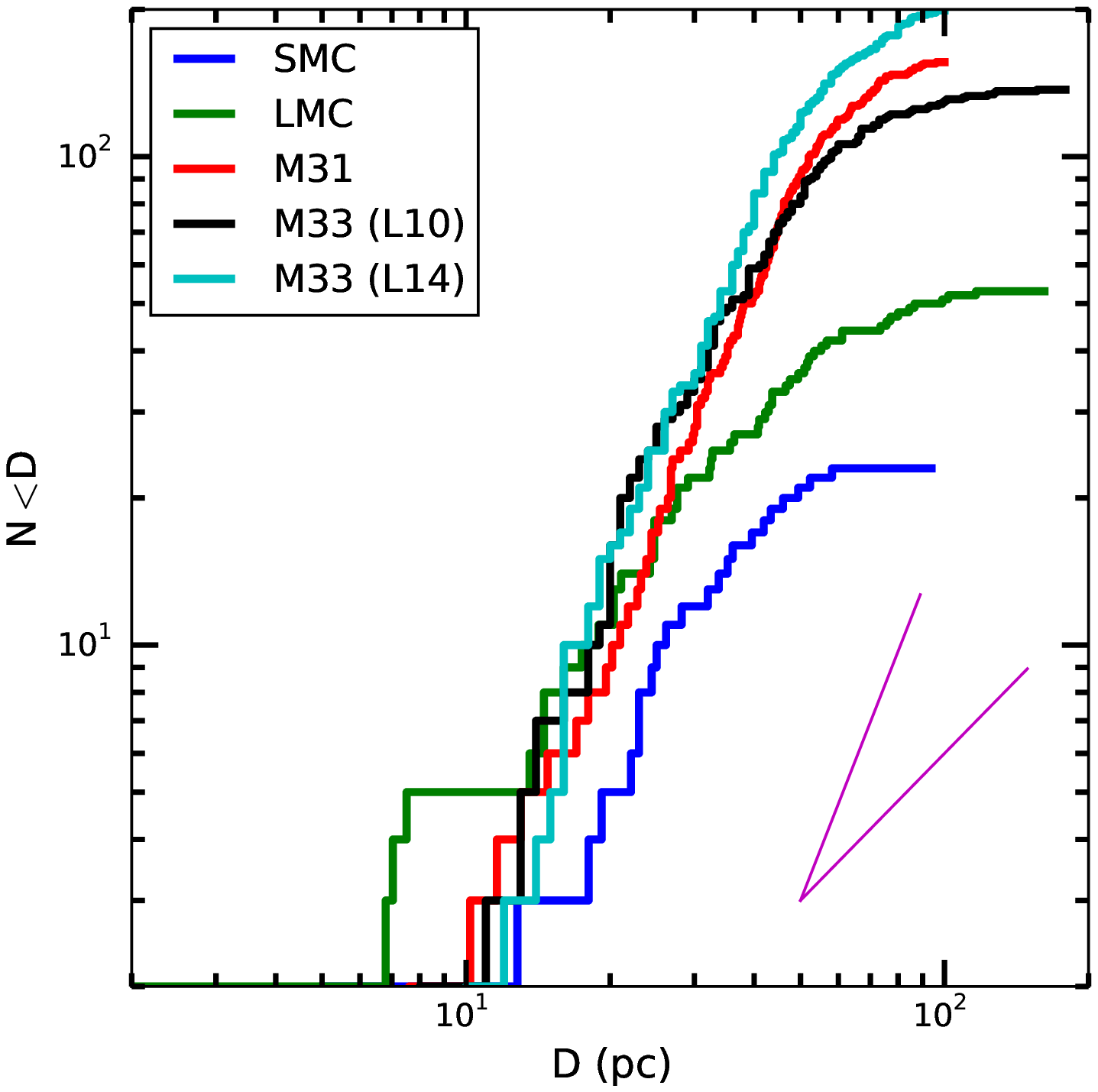}
\caption{The number of identified SNRs and SNR candidates in the Magellanic Clouds, M31 and M33 as a function of their diameter. The data for the SMC and LMC is from Badenes Maoz \& Draine \cite{badenes10}.  For M31, diameters were taken from the catalog of Lee \& Lee \cite{lee_m31}.  For M33, the catalogs of both Long \etal\ \cite{long10} and Lee \& Lee \cite{lee_m33} are shown.  Lines indicating the expected slopes of the curves for free expansion and Sedov evolution are also plotted.  Note that although the cumulative N$<$D - D  relation is shown here, the analysis of the distribution is more properly done on the differential number density, since the data points in the cumulative distribution are not statistically independent.}
\label{fig_number_diam}
\end{figure}

A natural question arising in any attempt to explain the properties of any SNR sample is how to explain the distribution of diameters in the sample.  If most SNRs  are in the Sedov phase, then one would naively expect the number of SNRs in a sample with diameters less than D (N$<D$) to increase as D$^{5/2}$. However, early versions of the N$<$D - D relationship for the Large Magellanic Clouds \cite{mathewson73,clarke76} and for M33 \cite{blair85} showed that the numbers increased roughly with D, as if SNRs expanded without significant deceleration to fairly large diameter.  The current N$<$D - D relationships for the Magellanic Clouds, M31 and M33 are shown in Fig.\ \ref{fig_number_diam}.  

The naive expectation that N should be proportional to  D$^{5/2}$ depends on whether a SNR sample is complete over a significant range of diameters.  Hughes, Helfand \& Kahn \cite{hughes84} showed that for reasonable variations in SN energy and ISM density, one could produce a flat   N$<$D - D  relation  if the sample was X-ray flux limited even if the majority of the SNRs were in the Sedov phase.  The reason for this is illustrated in the right hand panel of Fig.\ \ref{fig_xray_sensitivity}; a SNR expanding into a higher density ISM has  higher peak luminosities but fades away at lower diameters than a SNR from an identical SN expanding into a lower density medium.  As a result, in an X-ray flux limited sample, SNRs expanding into a lower density medium are visible at larger diameters. Although the number of SNRs known in the LMC has grown since the work of  Hughes, Helfand \& Kahn, the N$<$D - D relationship for the Large Magellanic Clouds remains relatively flat.  A recent analysis of the N$<$D - D  relationship was carried out by Badenes, Maoz \& Draine \cite{badenes10}  in the Magellanic Clouds.  They argue that the current sample is mostly complete, and that most of these SNRs are in Sedov phase, but that the relation is largely governed by variations  the density of the ISM. 

The observational situation in other galaxies is less clear.   Badenes, Maoz \& Draine \cite{badenes10} also analyzed the M33 SNR sample of Long \etal\ \cite{long10} and found it to resemble that of the Magellanic Clouds in the range of 10-30 pc, while Lee \& Lee \cite{lee_m33} using their sample of 199 SNRs and SNR candidates find a power slope of 2.4 in the range 17 to 50 pc,  identical to that expected for a pure Sedov expansion.  Lee \& Lee attribute this discrepancy to differences in the sample and also to differences in the diameter measurements.  Interestingly, Lee \& Lee \cite{lee_m31} also find that the size distribution of all of the SNRs they identified in M31 is consistent with a Sedov expansion law.  A thorough analysis, with a detailed assessment of completeness and contamination, of both the M33 and M31 samples would be desirable. 

\subsection{Radio Surface Brightness - Diameter Relationship $\Sigma$-D \index{SNR Radio Surface Brightness - Diameter Relation}}

\begin{figure}[b]
\includegraphics[scale=.38]{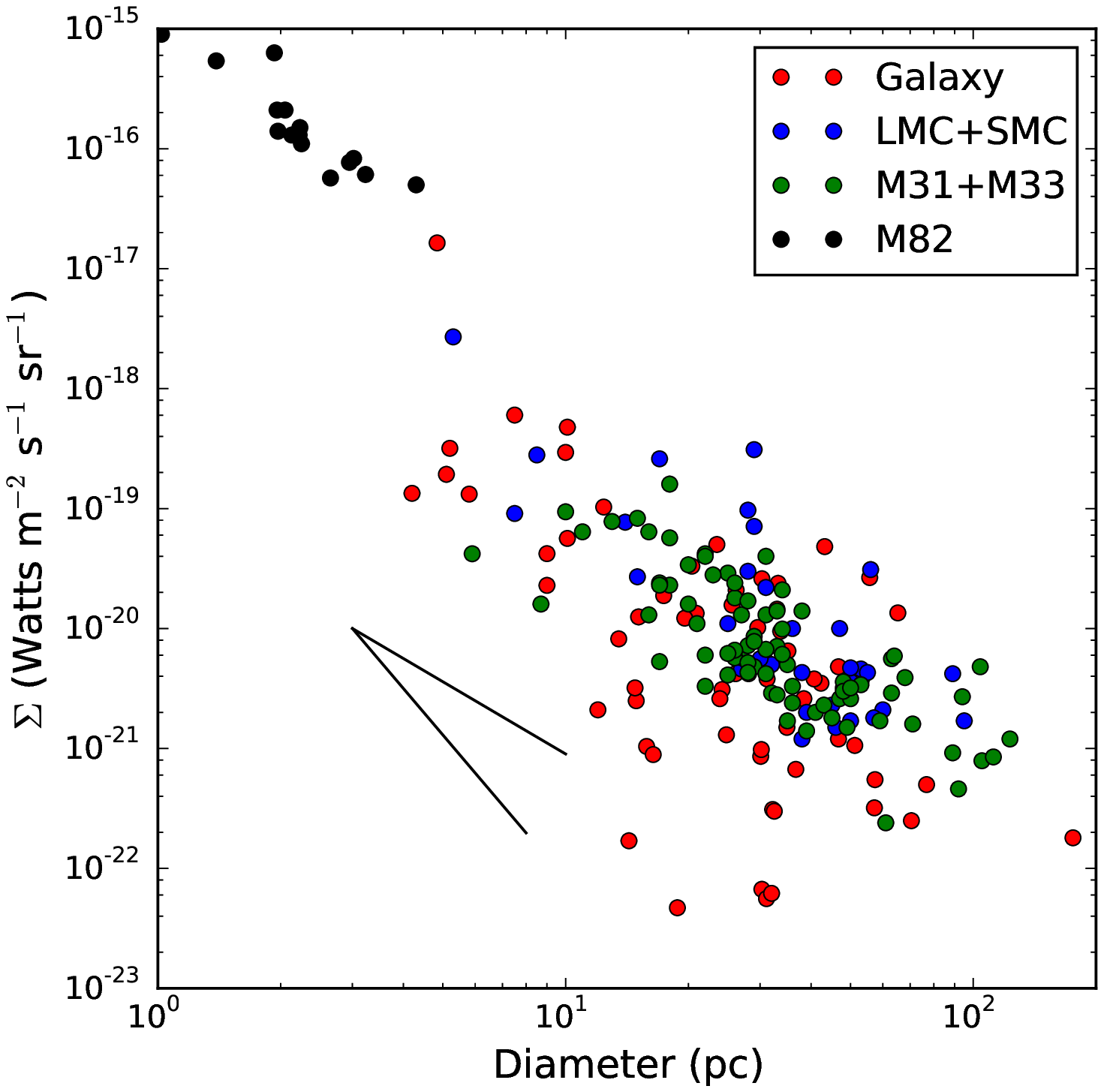}\includegraphics[scale=.38]{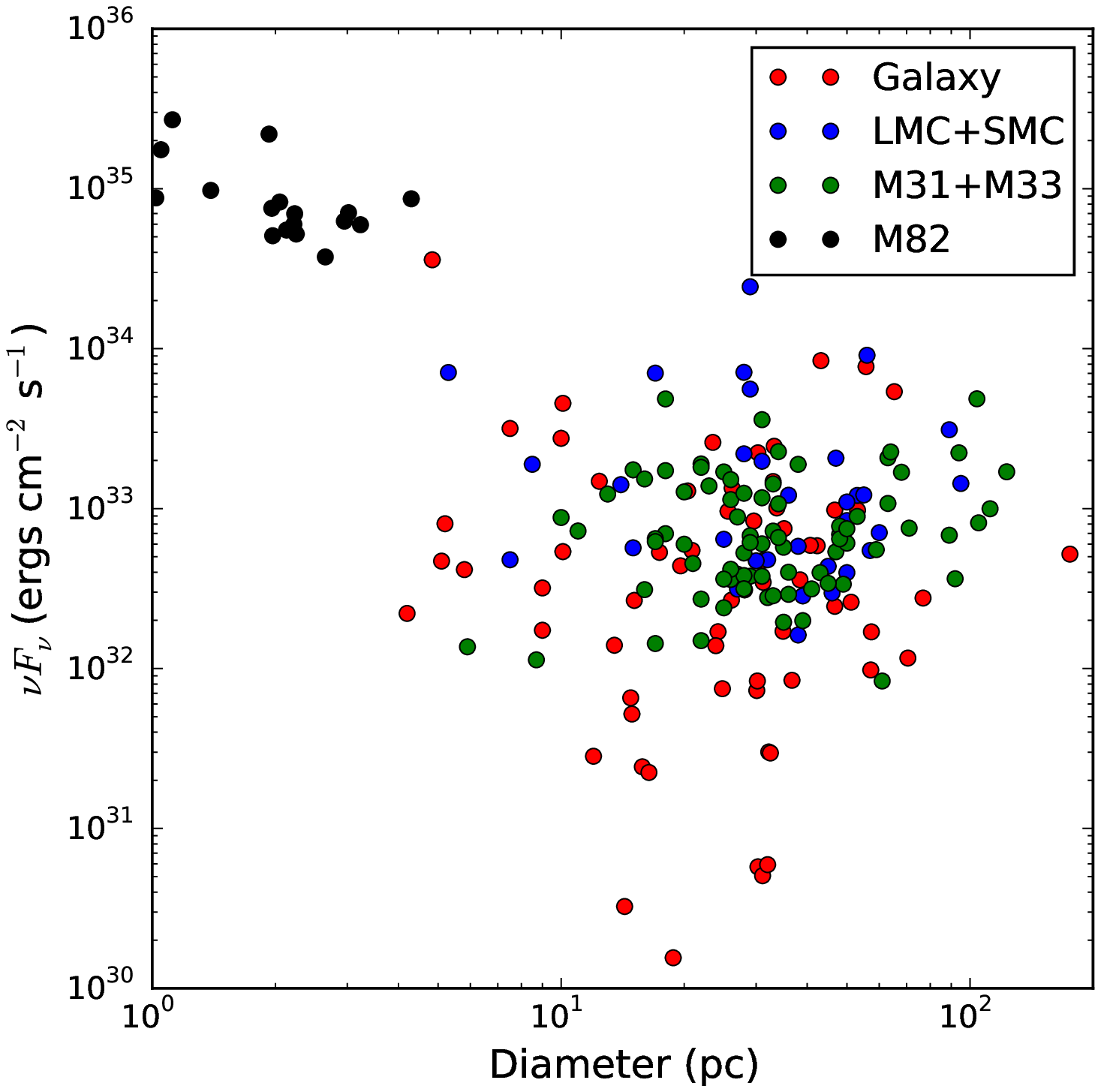}
\caption{Left: The radio surface brightness of  Galactic and extragalactic SNRs as a function of SNR diameter. The Galactic sample comprises 65 SNRs with distance estimates as compiled by Pavlovi{\'c} \etal\  \cite{pavlovic14}.  The extragalactic sample, divided into objects from the Magellanic Clouds, the Local Group galaxies M31 and M33, and M83 is taken from the compilation of {Uro{\v s}evi{\'c}} \etal\ \cite{urosevic05}. The data for M82 is from Huang \etal\ \cite{huang94}. The solid curves indicate power law slopes of -2 and -4. Right: The radio luminosity of  Galactic and extragalactic SNRs as a function of SNR diameter.  }
\label{fig_sigma_d}
\end{figure}

As noted earlier, the study of SNRs as a class of objects began with the identification of SNRs in the Galaxy and the Magellanic Clouds as extended radio sources.  The relationship between radio surface brightness and SNR diameter, the  so-called $\Sigma$ - D  relationship, was one of the first correlations to be identified in early SNR samples \cite{clark76}, and remains one whose importance continues to be discussed.  Small diameter SNRs typically have higher surface brightnesses than larger ones.  Sklovskii \cite{sklovskii60} pointed out that this fact, expected on theoretical grounds,  could be used to estimate distances to SNRs if it could be calibrated with SNRs of known distances. For Galactic SNRs, some of which have only been observed at radio wavelengths, it continues to be advocated by some for this purpose, if there is no alternative \cite{case98,pavlovic13,pavlovic14}.  

The radio surface brightness as a function of diameter for a collection of Galactic SNRs (with distance estimates) and of objects in the Magellanic Clouds, M31, M33, and M82 is shown in the left panel of Fig.\ \ref{fig_sigma_d}.   It illustrates both the fact that there is an apparent relationship between size and radio surface brightness and the problem, that the radio surface brightness at a particular diameter varies by large factors.  Clearly as pointed out by many,  distance estimates using a $\Sigma$ - D  relationship, any distance estimate based on this method will be crude at best.  

The $\Sigma$ - D relation is usually expressed  in terms of a power law of the form $\Sigma = A D^{\beta}$. If all SNRs had the same radio luminosity, then  $\Sigma \propto D^{-2}$.  Errors arise not only from the fact that SNRs from identical SNe evolving in different ISM may have different radio luminosities at the same diameter, but also from sample completeness biases, and in some cases, the way in which the power law index is derived from the observational data \cite{green14_distribution}.

The advantage of the Galactic sample is that one has more detailed information about the SNRs, and can weed out pulsar dominated SNRs, e. g. the Crab Nebula, which are not expected to follow the same relationship as the shell-like SNRs; the disadvantage is that the uncertainties in the distance measurements in the SNRs used in determining the $\Sigma$ - D are quite difficult to quantify.   Most estimates of the power law exponent in the Galactic sample fall in the range of  -4$\pm$1. 
 Clark \& Caswell \cite{clark76} in an early analysis of the Galactic sample derived a value of about -3, similar to the value Mathewson \& Clark \cite{mathewson73} had found for LMC SNRs.  Case \& Bhattacharya \cite{case98} found a value 2.64 for a sample of 37 SNRs, which included Cas A, although their analysis approach has been criticized by Green \cite{green14_distribution}. {Pavlovi{\'c} \etal\ \cite{pavlovic13,pavlovic14}, who attempt to account for the distance uncertainties for individual objects,  find steeper exponents of order -5.  

Extragalactic samples of SNRs are all at  known relative distances, but have the disadvantage that completeness and sample purity (especially beyond the Local Group)  are difficult to estimate.   One of the earliest attempts to study the $\Sigma$ - D relationship in galaxies beyond the Local Group was made by Berkhuijsen \cite{berkuijsen86} who used a sample of 86 SNRs from the Galaxy, LMC,  M33 to derive a $\Sigma$ - D relationship.  She found that her sample could be defined by a power law index of -3.5, but that the dispersion in surface brightness at any diameter was of order a factor of 10, something that is clear from the data plotted in Fig.\ \ref{fig_sigma_d}.  Huang \etal\ \cite{huang94} obtained measurements of a group of small diameter SNRs in star-burst galaxy M82, and also found a $\Sigma$ - D  power law index of -3.5$\pm$0.1.   However,  {Uro{\v s}evi{\'c}} \etal\ \cite{urosevic05} recently analyzed the data for 11 galaxies, and conclude that an exponent of -2 is consistent with the data for 10 of these galaxies.  If this is correct, then it simply says that the luminosity of a SNR does not vary strongly with size, and that factors other than size are more important in determining the radio luminosity.  To illustrate this, the radio luminosity of the same sample of objects is shown in the right panel of Fig.\ \ref{fig_sigma_d}.  The M82 SNRs are clearly brighter than those in the rest of the sample, but the dispersion in luminosity of the other samples at any diameter is so large that there are no obvious trends with diameter.

\begin{figure}[b]
\includegraphics[scale=.38]{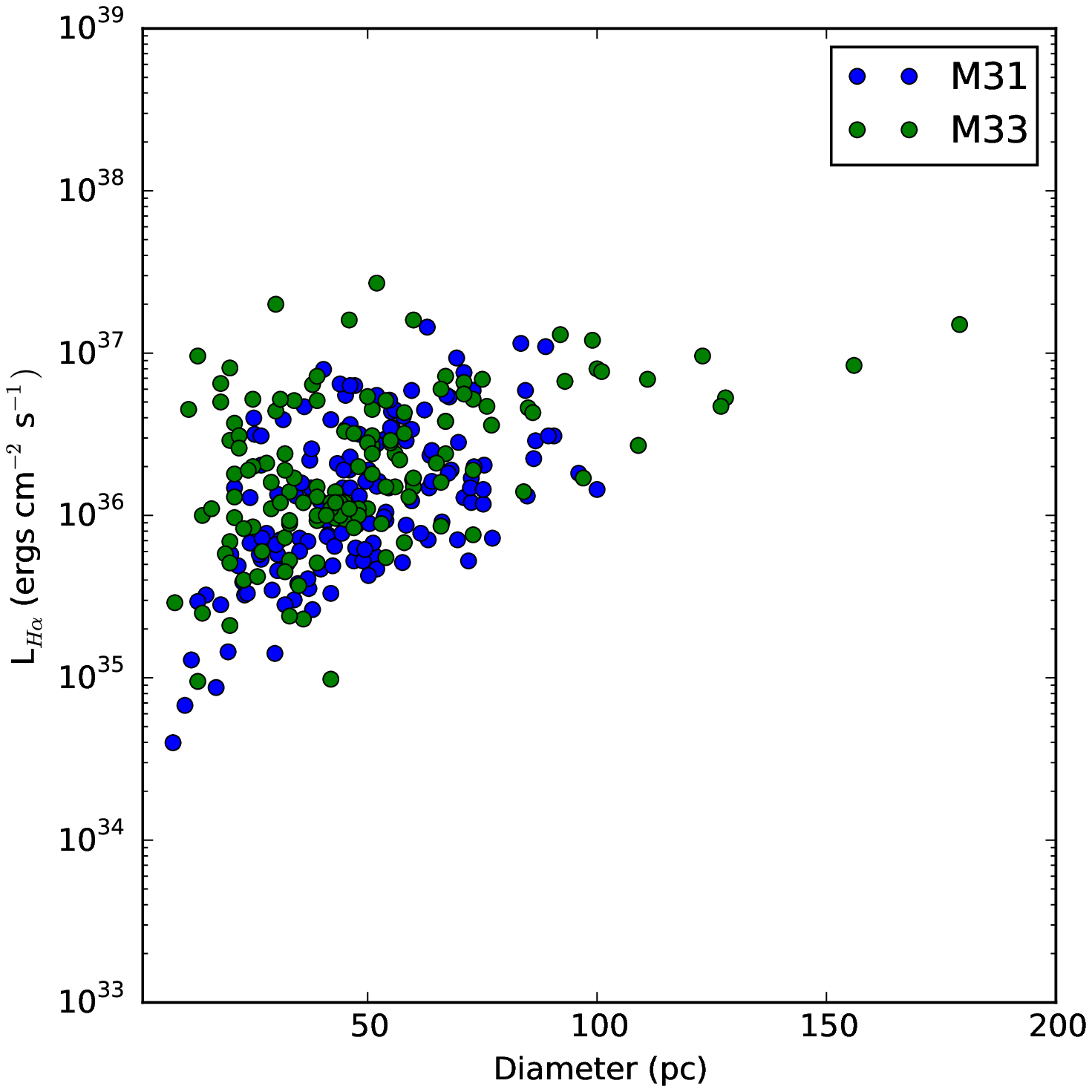}
\includegraphics[scale=.38]{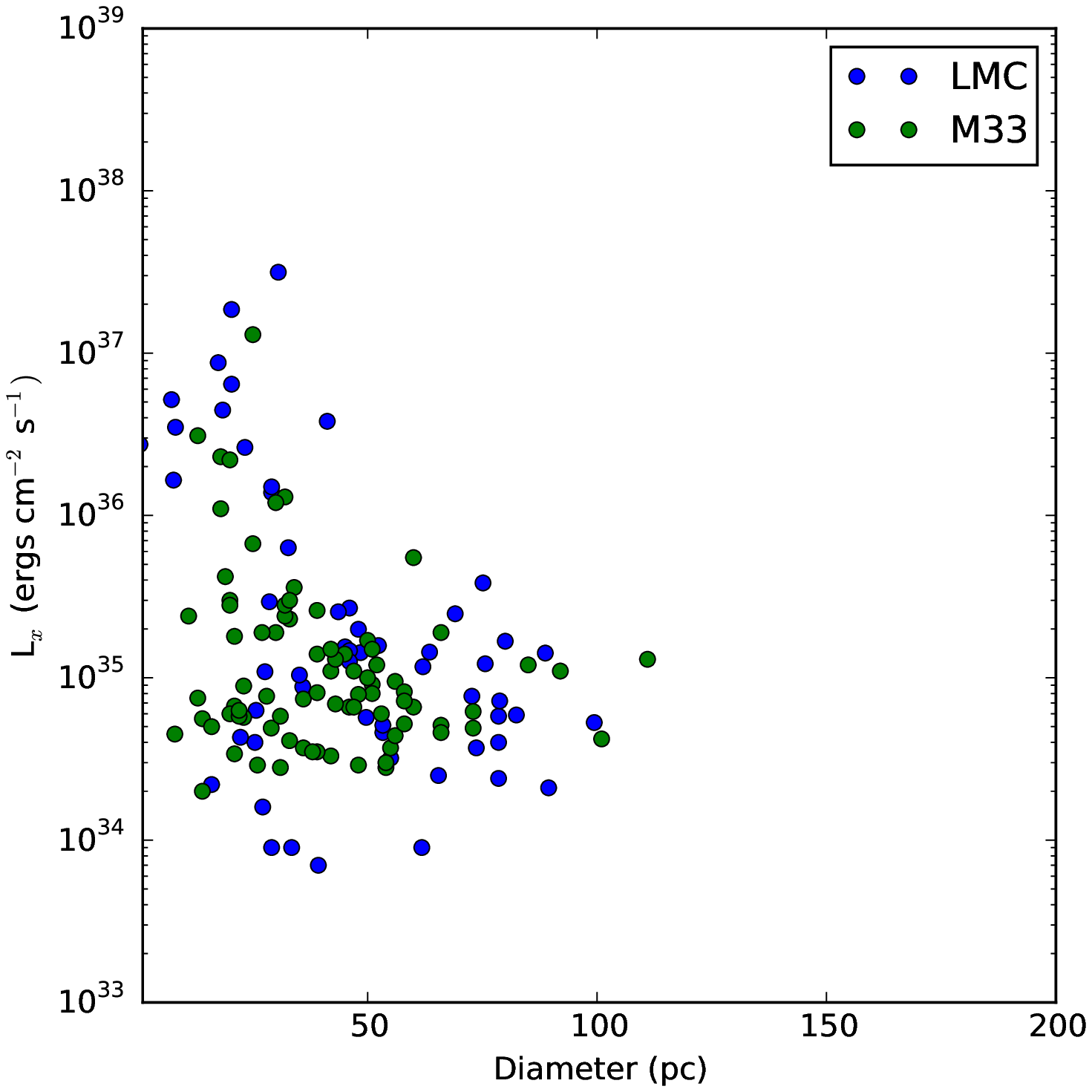}
\caption{Left:  The \HA\ luminosities of SNRs in M31 \cite{lee_m31} and M33 \cite{long10} as a function of diameter.  Right: The X-ray luminosities of SNRs in the LMC \cite{maggi15} and M33 \cite{long10}. Note that this is a semilog plot, unlike Fig. \ref{fig_sigma_d}. }
\label{fig_xlum_diam}
\end{figure}

\subsection{X-ray - Optical Comparisons \index{ SNRs: X-ray/Optical Comparisons} }

\begin{figure}[b]
\includegraphics[scale=.38]{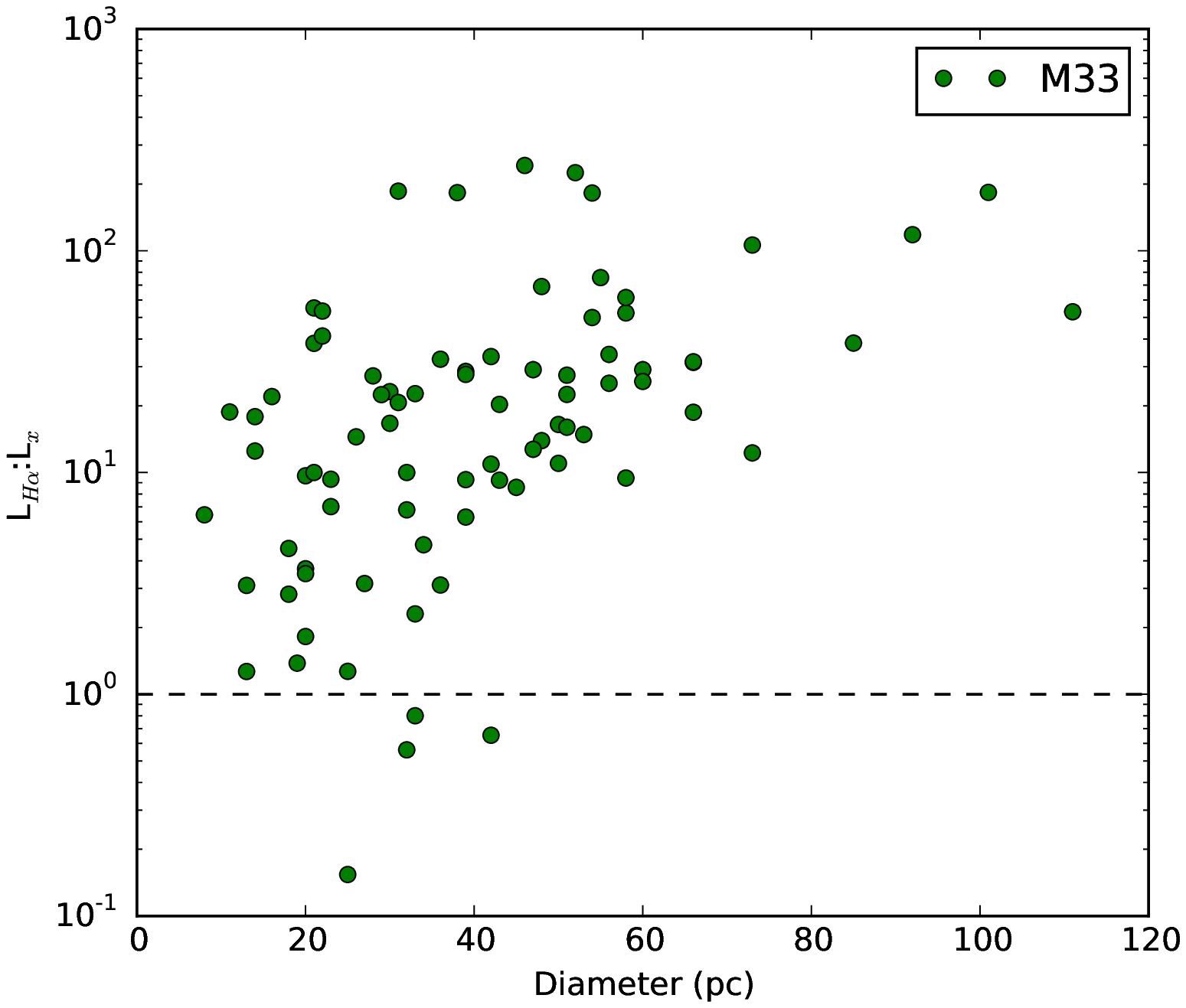}\includegraphics[scale=.38]{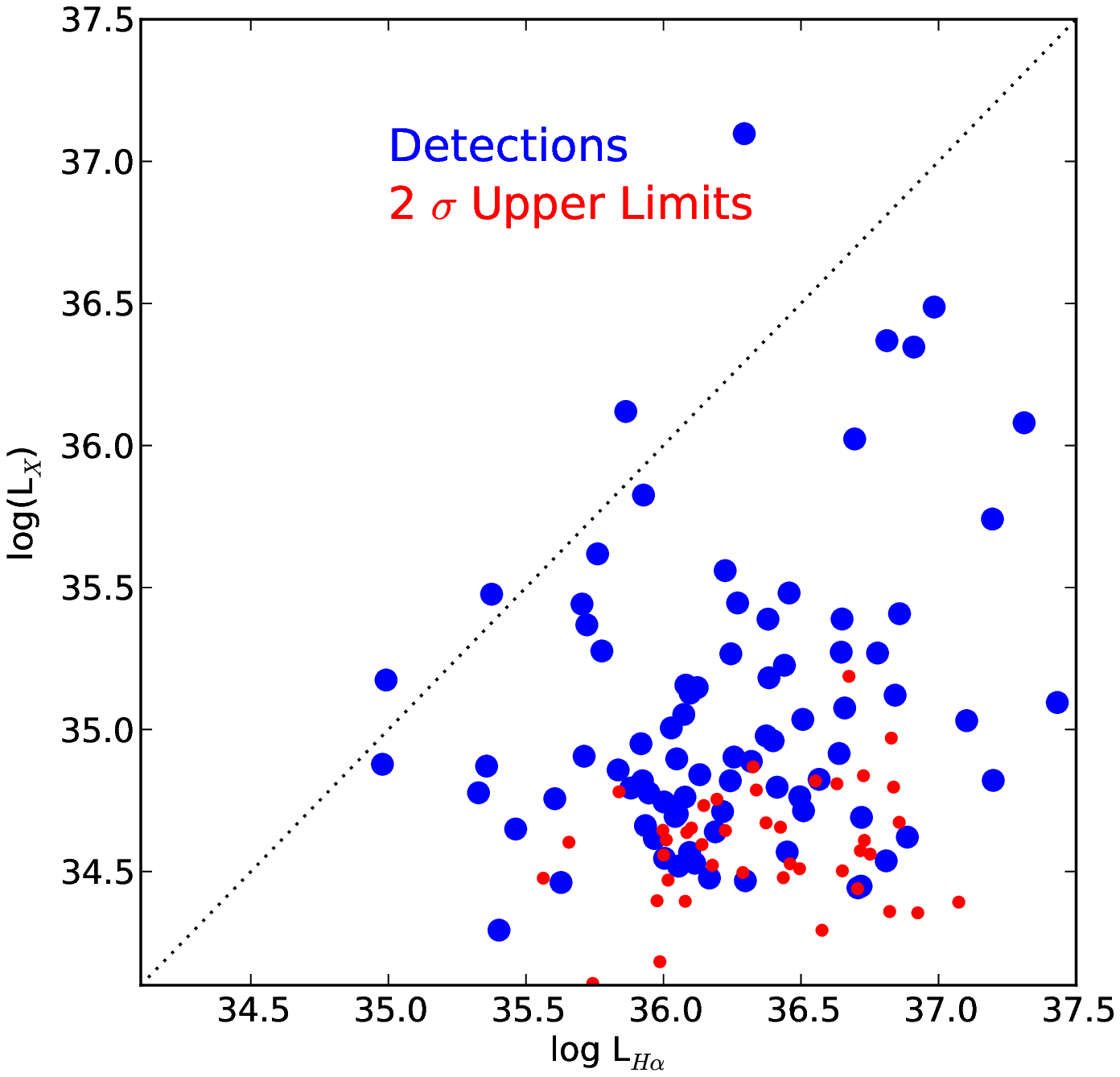}
\caption{Left: The ratio of \HA\ to X-ray luminosities of SNRs and SNR candidates in M33 as a function of SNR diameter, as described by Long \etal\ \cite{long10}.     Right: The \HA\ luminosities  as a function of X-ray luminosity, also from Long \etal\ \cite{long10}. The blue data points are for SNRs that were detected in X-rays; the red data points represent upper limits for the X-ray detection.}
\label{fig_ha}
\end{figure}

As we have discussed, optical emission from SNRs arises, in most cases, from radiative shocks driven into denser clouds in the ISM by the primary shock of a SN explosion, while X-ray emission arises from material heated by the primary shock.  As a result, one might guess that there would not be much correlation of \HA\ luminosity with diameter or with X-ray luminosity.  This is indeed the case.  The left panel of Fig.\ \ref{fig_xlum_diam} shows \HA\ luminosities of SNRs and SNR candidates in M31 and  M33 as a function of SNR diameter from the studies  of Lee \& Lee \cite{lee_m31} and Long \etal\ \cite{long10}, respectively.  The right panel of the same figure shows X-ray luminosities of SNRs in the LMC and M33 from the work of Maggi \etal\  \cite{maggi15} and Long \etal\ \cite{long10}.  In both cases, there is a lot of dispersion in luminosity at smaller diameters, but the amount of dispersion decreases towards higher diameters.  For the \HA\  images, the decrease in dispersion arises directly from the fact that optical searches for SNRs are surface brightness limited.  The X-ray surveys, by contrast, tend to be flux limited.  The upper envelopes of the two plots are not subject to selection effects.  Qualitatively at least, one can interpret the flatness that the \HA\ upper envelope of the \HA\ luminosity distribution as being due to the fact that the emission is arising from slower secondary shocks; the decline in the X-ray luminosities is due to the fact that at X-ray wavelengths, a SNR expanding into a denser medium cools to the point that the SNR is no longer detected even in soft X-rays. 

The \HA\ luminosities of SNRs are generally larger than their soft X-ray luminosities, as is indicated in Fig.\ \ref{fig_ha}, at least if simple count rate conversions are used to estimate X-ray luminosities.  This is not a selection effect; if SNRs existed that populated the high X-ray luminosity - low \HA\ luminosity section of the right panel of Fig.\ \ref{fig_ha}, they would have been detected as X-ray SNRs, even if they had not been identified as SNRs from their optical spectra.  There is clearly no correlation between X-ray and optical luminosity.

\subsection{The progenitors of  SNRs and the Type of the SN explosion \index{SNR progenitors}}

The most basic fact one would like to know about a SNR is whether the SNR is the product of Type Ia or core-collapse SN.   The best way to do this is to witness the SN and then to watch the development of the SNR as we are currently able to do with SN1987a, and a group of other less than 100 year old SNRs (see e. g. Milisavljevic \etal\ \cite{danny12}.  Second best, perhaps, is to obtain spectra of  the light echo of the SN explosion, which as has been done for one SN in the LMC \cite{rest08}, confirming it to be the result of a type Ia explosion. The others are also thought to be  of type Ia on the basis of their Balmer-dominated optical and X-ray Fe-rich spectra.  The mere detection of a light echo tends to favor a Ia explosion because core collapse SNe are generally fainter than Ia SNe.  }  However, it is much more common to attempt this determination from the properties of the SNR itself or by studies of the local stellar population.

Core collapse SNe ejecta are rich in CNO processed material, while ejecta from Ia SNe are rich in Fe, Si and other elements produced by explosive nucleosynthesis.  This material is heated as the reverse shock traverses the ejecta, and slowly mixes with shocked interstellar gas as the SNR expands.  It is more easily detectable in X-rays than at optical wavelengths since the material remains hot after the passage of the reverse shock and most optical emission is produced at the edge of the SNR in secondary shocks passing through material with interstellar abundances.  What is required are X-ay spectra with sufficient high signal to noise obtained with  high enough spectral resolution to carry out a detailed abundance analysis, and in some cases, enough spatial resolution to isolate ejecta material in the SNR interior from material emitting at the shock front.  With current instrumentation, this implies X-ray typing of SNRs is really only possible for SNRs in the Galaxy, in the Magellanic Clouds, and the brightest SNRs in M31 and M33.  

The first X-ray determinations of progenitor type in SNRs outside the galaxy were made by  Hughes \etal\ \cite{hughes95} who used \asca\  to show that two Balmer-dominated SNRs in the LMC were, as had been previously suggested by Tuohy \etal\ \cite{tuohy82}, the remnants of SNe of type Ia. As better X-ray spectra have been obtained, many more SNRs have been typed.  Maggi \etal\ recently concluded from an analysis of the XMM X-ray spectra that about 20 SNRs in the LMC can be classified as core-collapse or Ia SNRs on the basis of their X-ray spectra. and that six others can be classified (as core-collapse objects) on the basis of other properties, mainly related to the existence of a pulsar or pulsar wind nebula.  They use these results to argue that the ratio of CC:Ia SNRs is between 1.2 and 1.8, which is greater than the ratio of about 3:1 that Li \etal\ \cite{li11} measured from a volume limited sample of SNe or from abundances derived from studies of gas in galaxy clusters.  Although Maggi \etal\ consider the possibility that the unusual ratio of CC to Ia SNRs could be due  to multiple core-collapse SN exploding in the same star-formation region, they favor an explanation associated with the star formation history of the LMC.

The progenitors of a small number of SNRs can also be typed  on the basis of their optical spectra.  These include Cas A \cite{kirshner77}, G292+1.8 \cite{goss79} and Puppis A \cite{winkler85} in the Galaxy, and N132D \cite{morse95} in the LMC and E0102-72.3 \cite{blair00} in the SMC, all of which show spectra dominated by very strong oxygen lines, and many of which have other lines in their spectra, e.g. [Ne III], [Ne IV]  and [Ar IV], not normally seen in SNRs.  All of the young SNRs whose spectra are dominated by Balmer lines are most likely due to Ia explosions; a pre-requisite for a Balmer-dominated shock is a partially neutral ISM, and the UV emission  type II explosions is so strong that it should completely ionize the surrounding ISM \cite{tuohy82}.  (Portions of the  the Cygnus Loop, which is thought to be the result of a core-collapse explosion, are Balmer-dominated, but the majority of the filaments are radiative and this is a middle-aged SNR.)  

But SNRs which can be typed optically are rare. X-ray techniques are generally more powerful, and are likely to become even more powerful with the next generation of experiments, e. g.  Athena \cite{athena}, which will include non-dispersive, high resolution microcalorimeters  that will make elemental abundance determinations far more straightforward.

An alternative way to learn about the nature of the progenitor of a SNR is to study the underlying stellar population.  In the Magellanic Clouds, for example, Badenes \etal\  \cite{badenes09} has shown that all of the core-collapse SNRs are associated with regions of active star formation, while three out of the four Ia SNRs are associated with old metal poor populations.   Jennings \etal\ \cite{jennings12,jennings14} used HST data to construct color-magnitude diagrams of 33 M33 SNRs and 83 M31 SNRs and estimate the age and mass of the progenitor in each case.  They conclude that the distribution of masses is inconsistent with that expected from a Salpeter mass function; there are too few systems associated with very massive stars, suggesting an upper limit to the mass of stars that explode as SNe.

\section{Conclusion}

A great deal of progress in understanding SNRs and how they affect the Galaxy and the ISM has been made since  it was recognized about 50 years ago that SNRs constitute a large fraction of the brighter radio sources in the Galaxy.  Observationally, SNRs are very diverse, partly because the SNe that produces SNRs are diverse, but perhaps more  importantly because the circumstellar and interstellar environments into which SNRs explode is diverse. Large (but incomplete) samples of SNRs exist not only for the Galaxy but also for many nearby ($<$10 Mpc) galaxies.  Though most of the extragalactic SNRs have been discovered optically (through interference filter imaging which detects SNRs as emission nebulae with elevated [S II]:\HA\ ratios), it is important to continue to exploit techniques in other wavelength ranges -- radio, IR, X-ray, and $\gamma$-rays -- in the future.  Searches in multiple wavelength bands help to assure we have more complete SNR samples, and detection in multiple bands is the best way to be sure a SNR candidate is actually a SNR.   Studies in multiple bands are the best way to understand samples as a whole.  

Galactic samples are important, despite uncertainties in distance to individual objects and the effects of absorption in the plane; this is the only sample where we can study individual objects in detail and  study the oldest, faintest objects.  The extragalactic samples are important, despite our being able to learn less about individual objects,  because the samples in each galaxy are all at the same distance and can be observed and compared much more uniformly than the Galactic sample.  

There is still much to do in terms of interpretation of the existing samples. Because of the diversity of SNe and the environments in which they evolve, it is not straightforward interpret the existing data.  Although one can certainly expand the numbers of galaxies in which SNRs are identified in the future, the path to better understanding of SNRs as a class of objects is a more complete multiwavelength study of SNRs in the nearby galaxies with known SNR samples, especially at X-ray and radio wavelengths.

\begin{acknowledgement}  
Partial support for this work was provided by NASA through grants HST-GO-12462 and HST-GO-12762 issued by the Space Telescope Science Institute  and through grant  \chandra\ G0-13060  issued by  the \chandra\ X-ray Center. My understanding of SNRs as a class is largely due to conversations with my many collaborators, most notably William P. Blair and P. Frank Winkler.   
\end{acknowledgement}

\bibliography{snr}{}

\begin{thebibliography}{100}
\providecommand{\url}[1]{{#1}}
\providecommand{\urlprefix}{URL }
\expandafter\ifx\csname urlstyle\endcsname\relax
  \providecommand{\doi}[1]{DOI \discretionary{}{}{}#1}\else
  \providecommand{\doi}{DOI \discretionary{}{}{}\begingroup
  \urlstyle{rm}\Url}\fi

\bibitem{chu90}
Y.H. {Chu}, M.M. {Mac Low}, \apj \textbf{365}, 510 (1990).
\newblock \doi{10.1086/169505}

\bibitem{minkowski64}
R.~{Minkowski}, \araa \textbf{2}, 247 (1964).
\newblock \doi{10.1146/annurev.aa.02.090164.001335}

\bibitem{green14}
D.A. {Green}, Bulletin of the Astronomical Society of India \textbf{42}, 47
  (2014)

\bibitem{mathewson63}
D.S. {Mathewson}, J.R. {Healey}, \nat \textbf{199}, 681 (1963).
\newblock \doi{10.1038/199681a0}

\bibitem{mathewson73}
D.S. {Mathewson}, J.N. {Clarke}, \apj \textbf{180}, 725 (1973).
\newblock \doi{10.1086/152002}

\bibitem{maggi15}
P.~{Maggi}, F.~{Haberl}, P.J. {Kavanagh}, M.~{Sasaki}, L.M. {Bozzetto}, M.D.
  {Filipovi{\'c}}, G.~{Vasilopoulos}, W.~{Pietsch}, S.D. {Points}, Y.H. {Chu},
  J.~{Dickel}, M.~{Ehle}, R.~{Williams}, J.~{Greiner}, \aap \textbf{585}, A162
  (2016).
\newblock \doi{10.1051/0004-6361/201526932}

\bibitem{long10}
K.S. {Long}, W.P. {Blair}, P.F. {Winkler}, R.H. {Becker}, T.J. {Gaetz},
  P.~{Ghavamian}, D.J. {Helfand}, J.P. {Hughes}, R.P. {Kirshner}, K.D. {Kuntz},
  E.K. {McNeil}, T.G. {Pannuti}, P.P. {Plucinsky}, D.~{Saul},
  R.~{T{\"u}llmann}, B.~{Williams}, \apjs \textbf{187}, 495 (2010).
\newblock \doi{10.1088/0067-0049/187/2/495}

\bibitem{lee_m33}
J.H. {Lee}, M.G. {Lee}, \apj \textbf{793}, 134 (2014).
\newblock \doi{10.1088/0004-637X/793/2/134}

\bibitem{blair12}
W.P. {Blair}, P.F. {Winkler}, K.S. {Long}, \apjs \textbf{203}, 8 (2012).
\newblock \doi{10.1088/0067-0049/203/1/8}

\bibitem{blair15}
W.P. {Blair}, P.F. {Winkler}, K.S. {Long}, B.C. {Whitmore}, H.~{Kim},
  R.~{Soria}, K.D. {Kuntz}, P.P. {Plucinsky}, M.A. {Dopita}, C.~{Stockdale},
  \apj \textbf{800}, 118 (2015).
\newblock \doi{10.1088/0004-637X/800/2/118}

\bibitem{matonick97}
D.M. {Matonick}, R.A. {Fesen}, \apjs \textbf{112}, 49 (1997).
\newblock \doi{10.1086/313034}

\bibitem{vucetic15}
M.M. {Vu{\v c}eti{\'c}}, B.~{Arbutina}, D.~{Uro{\v s}evi{\'c}}, \mnras
  \textbf{446}, 943 (2015).
\newblock \doi{10.1093/mnras/stu2093}

\bibitem{krause08}
O.~{Krause}, S.M. {Birkmann}, T.~{Usuda}, T.~{Hattori}, M.~{Goto}, G.H.
  {Rieke}, K.A. {Misselt}, Science \textbf{320}, 1195 (2008).
\newblock \doi{10.1126/science.1155788}

\bibitem{fesen06}
R.A. {Fesen}, M.C. {Hammell}, J.~{Morse}, R.A. {Chevalier}, K.J. {Borkowski},
  M.A. {Dopita}, C.L. {Gerardy}, S.S. {Lawrence}, J.C. {Raymond}, S.~{van den
  Bergh}, \apj \textbf{645}, 283 (2006).
\newblock \doi{10.1086/504254}

\bibitem{delaney10}
T.~{DeLaney}, L.~{Rudnick}, M.D. {Stage}, J.D. {Smith}, K.~{Isensee}, J.~{Rho},
  G.E. {Allen}, H.~{Gomez}, T.~{Kozasa}, W.T. {Reach}, J.E. {Davis}, J.C.
  {Houck}, \apj \textbf{725}, 2038 (2010).
\newblock \doi{10.1088/0004-637X/725/2/2038}

\bibitem{grefenstette14}
B.W. {Grefenstette}, F.A. {Harrison}, S.E. {Boggs}, S.P. {Reynolds}, C.L.
  {Fryer}, K.K. {Madsen}, D.R. {Wik}, A.~{Zoglauer}, C.I. {Ellinger}, D.M.
  {Alexander}, H.~{An}, D.~{Barret}, F.E. {Christensen}, W.W. {Craig},
  K.~{Forster}, P.~{Giommi}, C.J. {Hailey}, A.~{Hornstrup}, V.M. {Kaspi},
  T.~{Kitaguchi}, J.E. {Koglin}, P.H. {Mao}, H.~{Miyasaka}, K.~{Mori},
  M.~{Perri}, M.J. {Pivovaroff}, S.~{Puccetti}, V.~{Rana}, D.~{Stern}, N.J.
  {Westergaard}, W.W. {Zhang}, \nat \textbf{506}, 339 (2014).
\newblock \doi{10.1038/nature12997}

\bibitem{koyama95}
K.~{Koyama}, R.~{Petre}, E.V. {Gotthelf}, U.~{Hwang}, M.~{Matsuura},
  M.~{Ozaki}, S.S. {Holt}, \nat \textbf{378}, 255 (1995).
\newblock \doi{10.1038/378255a0}

\bibitem{winkler14}
P.F. {Winkler}, B.J. {Williams}, S.P. {Reynolds}, R.~{Petre}, K.S. {Long},
  S.~{Katsuda}, U.~{Hwang}, \apj \textbf{781}, 65 (2014).
\newblock \doi{10.1088/0004-637X/781/2/65}

\bibitem{gaensler06}
B.M. {Gaensler}, P.O. {Slane}, \araa \textbf{44}, 17 (2006).
\newblock \doi{10.1146/annurev.astro.44.051905.092528}

\bibitem{park07}
S.~{Park}, J.P. {Hughes}, P.O. {Slane}, D.N. {Burrows}, B.M. {Gaensler},
  P.~{Ghavamian}, \apjl \textbf{670}, L121 (2007).
\newblock \doi{10.1086/524406}

\bibitem{haberl12}
F.~{Haberl}, R.~{Sturm}, J.~{Ballet}, D.J. {Bomans}, D.A.H. {Buckley}, M.J.
  {Coe}, R.~{Corbet}, M.~{Ehle}, M.D. {Filipovic}, M.~{Gilfanov},
  D.~{Hatzidimitriou}, N.~{La Palombara}, S.~{Mereghetti}, W.~{Pietsch},
  S.~{Snowden}, A.~{Tiengo}, \aap \textbf{545}, A128 (2012).
\newblock \doi{10.1051/0004-6361/201219758}

\bibitem{lee_m31}
J.H. {Lee}, M.G. {Lee}, \apj \textbf{786}, 130 (2014).
\newblock \doi{10.1088/0004-637X/786/2/130}

\bibitem{millar11}
W.C. {Millar}, G.L. {White}, M.D. {Filipovi{\'c}}, J.L. {Payne}, E.J.
  {Crawford}, T.G. {Pannuti}, W.D. {Staggs}, \apss \textbf{332}, 221 (2011).
\newblock \doi{10.1007/s10509-010-0556-y}

\bibitem{leonidaki13}
I.~{Leonidaki}, P.~{Boumis}, A.~{Zezas}, \mnras \textbf{429}, 189 (2013).
\newblock \doi{10.1093/mnras/sts324}

\bibitem{huang94}
Z.P. {Huang}, T.X. {Thuan}, R.A. {Chevalier}, J.J. {Condon}, Q.F. {Yin}, \apj
  \textbf{424}, 114 (1994).
\newblock \doi{10.1086/173876}

\bibitem{blair97}
W.P. {Blair}, K.S. {Long}, \apjs \textbf{108}, 261 (1997).
\newblock \doi{10.1086/312958}

\bibitem{blair14}
W.P. {Blair}, R.~{Chandar}, M.A. {Dopita}, P.~{Ghavamian}, D.~{Hammer}, K.D.
  {Kuntz}, K.S. {Long}, R.~{Soria}, B.C. {Whitmore}, P.F. {Winkler}, \apj
  \textbf{788}, 55 (2014).
\newblock \doi{10.1088/0004-637X/788/1/55}

\bibitem{franchetti12}
N.A. {Franchetti}, R.A. {Gruendl}, Y.H. {Chu}, B.C. {Dunne}, T.G. {Pannuti},
  K.D. {Kuntz}, C.H.R. {Chen}, C.K. {Grimes}, T.M. {Aldridge}, \aj
  \textbf{143}, 85 (2012).
\newblock \doi{10.1088/0004-6256/143/4/85}

\bibitem{sonbas10}
E.~{Sonba{\c s}}, A.~{Aky{\"u}z}, {\c S}.~{Balman}, M.E. {{\"O}zel}, \aap
  \textbf{517}, A91 (2010).
\newblock \doi{10.1051/0004-6361/200913858}

\bibitem{sonbas09}
E.~{Sonbas}, A.~{Akyuz}, S.~{Balman}, \aap \textbf{493}, 1061 (2009).
\newblock \doi{10.1051/0004-6361:200810690}

\bibitem{massey06}
P.~{Massey}, K.A.G. {Olsen}, P.W. {Hodge}, S.B. {Strong}, G.H. {Jacoby},
  W.~{Schlingman}, R.C. {Smith}, \aj \textbf{131}, 2478 (2006).
\newblock \doi{10.1086/503256}

\bibitem{blair04}
W.P. {Blair}, K.S. {Long}, \apjs \textbf{155}, 101 (2004).
\newblock \doi{10.1086/423958}

\bibitem{dopita77}
M.A. {Dopita}, \apjs \textbf{33}, 437 (1977).
\newblock \doi{10.1086/190435}

\bibitem{raymond79}
J.C. {Raymond}, \apjs \textbf{39}, 1 (1979).
\newblock \doi{10.1086/190562}

\bibitem{allen08}
M.G. {Allen}, B.A. {Groves}, M.A. {Dopita}, R.S. {Sutherland}, L.J. {Kewley},
  \apjs \textbf{178}, 20 (2008).
\newblock \doi{10.1086/589652}

\bibitem{chevalier78}
R.A. {Chevalier}, J.C. {Raymond}, \apjl \textbf{225}, L27 (1978).
\newblock \doi{10.1086/182785}

\bibitem{raymond10}
J.C. {Raymond}, P.F. {Winkler}, W.P. {Blair}, J.J. {Lee}, S.~{Park}, \apj
  \textbf{712}, 901 (2010).
\newblock \doi{10.1088/0004-637X/712/2/901}

\bibitem{tuohy82}
I.R. {Tuohy}, M.A. {Dopita}, D.S. {Mathewson}, K.S. {Long}, D.J. {Helfand},
  \apj \textbf{261}, 473 (1982).
\newblock \doi{10.1086/160358}

\bibitem{long81}
K.S. {Long}, D.J. {Helfand}, D.A. {Grabelsky}, \apj \textbf{248}, 925 (1981).
\newblock \doi{10.1086/159222}

\bibitem{fesen15}
R.A. {Fesen}, J.M.M. {Neustadt}, C.S. {Black}, A.H.D. {Koeppel}, \apj
  \textbf{812}, 37 (2015).
\newblock \doi{10.1088/0004-637X/812/1/37}

\bibitem{kirshner77}
R.P. {Kirshner}, R.A. {Chevalier}, \apj \textbf{218}, 142 (1977).
\newblock \doi{10.1086/155666}

\bibitem{goss79}
W.M. {Goss}, P.A. {Shaver}, W.J. {Zealey}, P.~{Murdin}, D.H. {Clark}, \mnras
  \textbf{188}, 357 (1979).
\newblock \doi{10.1093/mnras/188.2.357}

\bibitem{dopita84}
M.A. {Dopita}, L.~{Binette}, I.R. {Tuohy}, \apj \textbf{282}, 142 (1984).
\newblock \doi{10.1086/162185}

\bibitem{finkelstein06}
S.L. {Finkelstein}, J.A. {Morse}, J.C. {Green}, J.L. {Linsky}, J.M. {Shull},
  T.P. {Snow}, J.T. {Stocke}, K.R. {Brownsberger}, D.C. {Ebbets},
  E.~{Wilkinson}, S.R. {Heap}, C.~{Leitherer}, B.D. {Savage}, O.H. {Siegmund},
  A.~{Stern}, \apj \textbf{641}, 919 (2006).
\newblock \doi{10.1086/500570}

\bibitem{long89}
K.S. {Long}, W.P. {Blair}, W.~{Krzeminski}, \apjl \textbf{340}, L25 (1989).
\newblock \doi{10.1086/185430}

\bibitem{kirshner80}
R.P. {Kirshner}, W.P. {Blair}, \apj \textbf{236}, 135 (1980).
\newblock \doi{10.1086/157726}

\bibitem{long12}
K.S. {Long}, W.P. {Blair}, L.E.H. {Godfrey}, K.D. {Kuntz}, P.P. {Plucinsky},
  R.~{Soria}, C.J. {Stockdale}, B.C. {Whitmore}, P.F. {Winkler}, \apj
  \textbf{756}, 18 (2012).
\newblock \doi{10.1088/0004-637X/756/1/18}

\bibitem{dubner15}
G.~{Dubner}, E.~{Giacani}, \aapr \textbf{23}, 3 (2015).
\newblock \doi{10.1007/s00159-015-0083-5}

\bibitem{kargaltsev15}
O.~{Kargaltsev}, B.~{Cerutti}, Y.~{Lyubarsky}, E.~{Striani}, \ssr  (2015).
\newblock \doi{10.1007/s11214-015-0171-x}

\bibitem{filipovic05}
M.D. {Filipovi{\'c}}, J.L. {Payne}, W.~{Reid}, C.W. {Danforth},
  L.~{Staveley-Smith}, P.A. {Jones}, G.L. {White}, \mnras \textbf{364}, 217
  (2005).
\newblock \doi{10.1111/j.1365-2966.2005.09554.x}

\bibitem{gordon99}
S.M. {Gordon}, N.~{Duric}, R.P. {Kirshner}, W.M. {Goss}, F.~{Viallefond}, \apjs
  \textbf{120}, 247 (1999).
\newblock \doi{10.1086/313175}

\bibitem{lacey01}
C.K. {Lacey}, N.~{Duric}, \apj \textbf{560}, 719 (2001).
\newblock \doi{10.1086/323048}

\bibitem{chomiuk_search}
L.~{Chomiuk}, E.M. {Wilcots}, \aj \textbf{137}, 3869 (2009).
\newblock \doi{10.1088/0004-6256/137/4/3869}

\bibitem{pfeffermann96}
E.~{Pfeffermann}, B.~{Aschenbach}, in \emph{Roentgenstrahlung from the
  Universe}, ed. by H.U. {Zimmermann}, J.~{Tr{\"u}mper}, H.~{Yorke} (1996), pp.
  267--268

\bibitem{aschenbach98}
B.~{Aschenbach}, \nat \textbf{396}, 141 (1998).
\newblock \doi{10.1038/24103}

\bibitem{xspec}
K.A. {Arnaud}, in \emph{Astronomical Data Analysis Software and Systems V},
  \emph{Astronomical Society of the Pacific Conference Series}, vol. 101, ed.
  by G.H. {Jacoby}, J.~{Barnes} (1996), \emph{Astronomical Society of the
  Pacific Conference Series}, vol. 101, p.~17

\bibitem{schaudel02}
D.~{Schaudel}, W.~{Becker}, W.~{Voges}, B.~{Aschenbach}, W.~{Reich},
  M.~{Weisskopf}, in \emph{Neutron Stars in Supernova Remnants},
  \emph{Astronomical Society of the Pacific Conference Series}, vol. 271, ed.
  by P.O. {Slane}, B.M. {Gaensler} (2002), \emph{Astronomical Society of the
  Pacific Conference Series}, vol. 271, p. 391

\bibitem{huang14}
R.H.H. {Huang}, J.H.K. {Wu}, C.Y. {Hui}, K.A. {Seo}, L.~{Trepl}, A.K.H. {Kong},
  \apj \textbf{785}, 118 (2014).
\newblock \doi{10.1088/0004-637X/785/2/118}

\bibitem{robbins12}
W.J. {Robbins}, B.M. {Gaensler}, T.~{Murphy}, S.~{Reeves}, A.J. {Green}, \mnras
  \textbf{419}, 2623 (2012).
\newblock \doi{10.1111/j.1365-2966.2011.19912.x}

\bibitem{busser96}
J.U. {Busser}, R.~{Egger}, B.~{Aschenbach}, \aap \textbf{310}, L1 (1996)

\bibitem{hui12}
C.Y. {Hui}, K.A. {Seo}, R.H.H. {Huang}, L.~{Trepl}, Y.J. {Woo}, T.N. {Lu},
  A.K.H. {Kong}, F.M. {Walter}, \apj \textbf{750}, 7 (2012).
\newblock \doi{10.1088/0004-637X/750/1/7}

\bibitem{erosita14}
P.~{Predehl}, R.~{Andritschke}, W.~{Becker}, W.~{Bornemann},
  H.~{Br{\"a}uninger}, H.~{Brunner}, T.~{Boller}, V.~{Burwitz}, W.~{Burkert},
  N.~{Clerc}, E.~{Churazov}, D.~{Coutinho}, K.~{Dennerl}, J.~{Eder},
  V.~{Emberger}, T.~{Eraerds}, M.J. {Freyberg}, P.~{Friedrich},
  M.~{F{\"u}rmetz}, A.~{Georgakakis}, C.~{Grossberger}, F.~{Haberl},
  O.~{H{\"a}lker}, G.~{Hartner}, G.~{Hasinger}, J.~{Hoelzl}, H.~{Huber},
  A.~{von Kienlin}, W.~{Kink}, I.~{Kreykenbohm}, G.~{Lamer}, I.~{Lomakin},
  I.~{Lapchov}, L.~{Lovisari}, N.~{Meidinger}, A.~{Merloni}, B.~{Mican},
  J.~{Mohr}, S.~{M{\"u}ller}, K.~{Nandra}, F.~{Pacaud}, M.N. {Pavlinsky},
  E.~{Perinati}, E.~{Pfeffermann}, D.~{Pietschner}, J.~{Reiffers},
  T.~{Reiprich}, J.~{Robrade}, M.~{Salvato}, A.E. {Santangelo}, M.~{Sasaki},
  H.~{Scheuerle}, C.~{Schmid}, J.~{Schmitt}, A.D. {Schwope}, R.~{Sunyaev},
  C.~{Tenzer}, L.~{Tiedemann}, W.~{Xu}, V.~{Yaroshenko}, S.~{Walther},
  M.~{Wille}, J.~{Wilms}, Y.Y. {Zhang}, in \emph{Society of Photo-Optical
  Instrumentation Engineers (SPIE) Conference Series}, \emph{Society of
  Photo-Optical Instrumentation Engineers (SPIE) Conference Series}, vol. 9144
  (2014), \emph{Society of Photo-Optical Instrumentation Engineers (SPIE)
  Conference Series}, vol. 9144, p. 91441T.
\newblock \doi{10.1117/12.2055426}

\bibitem{kahabka97}
P.~{Kahabka}, E.P.J. {van den Heuvel}, \araa \textbf{35}, 69 (1997).
\newblock \doi{10.1146/annurev.astro.35.1.69}

\bibitem{stiele11}
H.~{Stiele}, W.~{Pietsch}, F.~{Haberl}, D.~{Hatzidimitriou}, R.~{Barnard}, B.F.
  {Williams}, A.K.H. {Kong}, U.~{Kolb}, \aap \textbf{534}, A55 (2011).
\newblock \doi{10.1051/0004-6361/201015270}

\bibitem{leonidaki10}
I.~{Leonidaki}, A.~{Zezas}, P.~{Boumis}, \apj \textbf{725}, 842 (2010).
\newblock \doi{10.1088/0004-637X/725/1/842}

\bibitem{mouri00}
H.~{Mouri}, K.~{Kawara}, Y.~{Taniguchi}, \apj \textbf{528}, 186 (2000).
\newblock \doi{10.1086/308142}

\bibitem{oliva89}
E.~{Oliva}, A.F.M. {Moorwood}, I.J. {Danziger}, \aap \textbf{214}, 307 (1989)

\bibitem{oliva90}
E.~{Oliva}, A.F.M. {Moorwood}, I.J. {Danziger}, \aap \textbf{240}, 453 (1990)

\bibitem{greenhouse97}
M.A. {Greenhouse}, S.~{Satyapal}, C.E. {Woodward}, J.~{Fischer}, K.L.
  {Thompson}, W.J. {Forrest}, J.L. {Pipher}, N.~{Raines}, H.A. {Smith}, D.M.
  {Watson}, R.J. {Rudy}, \apj \textbf{476}, 105 (1997)

\bibitem{morel02}
T.~{Morel}, R.~{Doyon}, N.~{St-Louis}, \mnras \textbf{329}, 398 (2002).
\newblock \doi{10.1046/j.1365-8711.2002.05026.x}

\bibitem{lee14}
J.J. {Lee}, B.C. {Koo}, Y.H. {Lee}, H.G. {Lee}, J.H. {Shinn}, H.J. {Kim},
  Y.~{Kim}, T.S. {Pyo}, D.S. {Moon}, S.C. {Yoon}, M.Y. {Chun}, D.~{Froebrich},
  C.J. {Davis}, W.P. {Varricatt}, J.~{Kyeong}, N.~{Hwang}, B.G. {Park}, M.G.
  {Lee}, H.M. {Lee}, M.~{Ishiguro}, \mnras \textbf{443}, 2650 (2014).
\newblock \doi{10.1093/mnras/stu1146}

\bibitem{rosenberg12}
M.J.F. {Rosenberg}, P.P. {van der Werf}, F.P. {Israel}, \aap \textbf{540}, A116
  (2012).
\newblock \doi{10.1051/0004-6361/201218772}

\bibitem{long14}
K.S. {Long}, K.D. {Kuntz}, W.P. {Blair}, L.~{Godfrey}, P.P. {Plucinsky},
  R.~{Soria}, C.~{Stockdale}, P.F. {Winkler}, \apjs \textbf{212}, 21 (2014).
\newblock \doi{10.1088/0067-0049/212/2/21}

\bibitem{arendt89}
R.G. {Arendt}, \apjs \textbf{70}, 181 (1989).
\newblock \doi{10.1086/191337}

\bibitem{saken92}
J.M. {Saken}, R.A. {Fesen}, J.M. {Shull}, \apjs \textbf{81}, 715 (1992).
\newblock \doi{10.1086/191703}

\bibitem{pinheiro11}
D.~{Pinheiro Gon{\c c}alves}, A.~{Noriega-Crespo}, R.~{Paladini}, P.G.
  {Martin}, S.J. {Carey}, \aj \textbf{142}, 47 (2011).
\newblock \doi{10.1088/0004-6256/142/2/47}

\bibitem{reach06}
W.T. {Reach}, J.~{Rho}, A.~{Tappe}, T.G. {Pannuti}, C.L. {Brogan}, E.B.
  {Churchwell}, M.R. {Meade}, B.~{Babler}, R.~{Indebetouw}, B.A. {Whitney}, \aj
  \textbf{131}, 1479 (2006).
\newblock \doi{10.1086/499306}

\bibitem{acero15}
F.~{Acero}, M.~{Ackermann}, M.~{Ajello}, L.~{Baldini}, J.~{Ballet},
  G.~{Barbiellini}, D.~{Bastieri}, R.~{Bellazzini}, E.~{Bissaldi},
  R.~{Blandford}, E.D. {Bloom}, R.~{Bonino}, E.~{Bottacini}, J.~{Bregeon},
  P.~{Bruel}, R.~{Buehler}, S.~{Buson}, G.A. {Caliandro}, R.A. {Cameron},
  R.~{Caputo}, M.~{Caragiulo}, P.A. {Caraveo}, J.M. {Casandjian},
  E.~{Cavazzuti}, C.~{Cecchi}, A.~{Chekhtman}, J.~{Chiang}, G.~{Chiaro},
  S.~{Ciprini}, R.~{Claus}, J.M. {Cohen}, J.~{Cohen-Tanugi}, L.R. {Cominsky},
  B.~{Condon}, J.~{Conrad}, S.~{Cutini}, F.~{D'Ammando}, A.~{Angelis},
  F.~{Palma}, R.~{Desiante}, S.W. {Digel}, L.~{Venere}, P.S. {Drell},
  A.~{Drlica-Wagner}, C.~{Favuzzi}, E.C. {Ferrara}, A.~{Franckowiak},
  Y.~{Fukazawa}, Prof., S.~{Funk}, Prof., P.~{Fusco}, F.~{Gargano},
  D.~{Gasparrini}, N.~{Giglietto}, P.~{Giommi}, F.~{Giordano}, M.~{Giroletti},
  T.~{Glanzman}, G.~{Godfrey}, G.~{Gomez-Vargas}, I.A. {Grenier}, M.H.
  {Grondin}, L.~{Guillemot}, S.~{Guiriec}, M.~{Gustafsson}, D.~{Hadasch}, A.K.
  {Harding}, M.~{Hayashida}, E.~{Hays}, J.W. {Hewitt}, A.B. {Hill}, D.~{Horan},
  X.~{Hou}, G.~{Iafrate}, T.~{Jogler}, G.~{J'ohannesson}, A.S. {Johnson},
  T.~{Kamae}, H.~{Katagiri}, J.~{Kataoka}, Prof., J.~{Katsuta}, M.~{Kerr},
  J.~{Knodlseder}, D.~{Kocevski}, Prof., M.~{Kuss}, H.~{Laffon}, J.~{Lande},
  S.~{Larsson}, L.~{Latronico}, M.~{Lemoine-Goumard}, J.~{Li}, L.~{Li},
  F.~{Longo}, F.~{Loparco}, M.N. {Lovellette}, P.~{Lubrano}, J.~{Magill},
  S.~{Maldera}, M.~{Marelli}, M.~{Mayer}, M.N. {Mazziotta}, P.F. {Michelson},
  W.~{Mitthumsiri}, T.~{Mizuno}, A.A. {Moiseev}, M.E. {Monzani}, E.~{Moretti},
  A.~{Morselli}, I.V. {Moskalenko}, S.~{Murgia}, Prof., R.~{Nemmen}, Prof.,
  E.~{Nuss}, T.~{Ohsugi}, N.~{Omodei}, M.~{Orienti}, E.~{Orlando}, J.F.
  {Ormes}, D.~{Paneque}, J.S. {Perkins}, M.~{Pesce-Rollins}, V.~{Petrosian},
  Prof., F.~{Piron}, G.~{Pivato}, T.~{Porter}, S.~{Rain`o}, R.~{Rando},
  M.~{Razzano}, S.~{Razzaque}, A.~{Reimer}, O.~{Reimer}, Prof., M.~{Renaud},
  T.~{Reposeur}, M.~{Romain Rousseau}, P.M. {Parkinson}, J.~{Schmid},
  A.~{Schulz}, C.~{Sgr`o}, E.J. {Siskind}, F.~{Spada}, G.~{Spandre},
  P.~{Spinelli}, A.W. {Strong}, D.~{Suson}, H.~{Tajima}, H.~{Takahashi},
  T.~{Tanaka}, J.B. {Thayer}, D.J. {Thompson}, L.~{Tibaldo}, O.~{Tibolla}, D.F.
  {Torres}, Prof., G.~{Tosti}, E.~{Troja}, Y.~{Uchiyama}, G.~{Vianello},
  B.~{Wells}, K.~{Wood}, M.~{Wood}, M.~{Yassine}, S.~{Zimmer}, ArXiv e-prints
  (2015)

\bibitem{green14_distribution}
D.A. {Green}, in \emph{IAU Symposium}, \emph{IAU Symposium}, vol. 296, ed. by
  A.~{Ray}, R.A. {McCray} (2014), \emph{IAU Symposium}, vol. 296, pp. 188--196.
\newblock \doi{10.1017/S1743921313009459}

\bibitem{IPHAS}
J.E. {Drew}, R.~{Greimel}, M.J. {Irwin}, A.~{Aungwerojwit}, M.J. {Barlow},
  R.L.M. {Corradi}, J.J. {Drake}, B.T. {G{\"a}nsicke}, P.~{Groot}, A.~{Hales},
  E.C. {Hopewell}, J.~{Irwin}, C.~{Knigge}, P.~{Leisy}, D.J. {Lennon},
  A.~{Mampaso}, M.R.W. {Masheder}, M.~{Matsuura}, L.~{Morales-Rueda}, R.A.H.
  {Morris}, Q.A. {Parker}, S.~{Phillipps}, P.~{Rodriguez-Gil}, G.~{Roelofs},
  I.~{Skillen}, J.L. {Sokoloski}, D.~{Steeghs}, Y.C. {Unruh}, K.~{Viironen},
  J.S. {Vink}, N.A. {Walton}, A.~{Witham}, N.~{Wright}, A.A. {Zijlstra},
  A.~{Zurita}, \mnras \textbf{362}, 753 (2005).
\newblock \doi{10.1111/j.1365-2966.2005.09330.x}

\bibitem{sabin13}
L.~{Sabin}, Q.A. {Parker}, M.E. {Contreras}, L.~{Olgu{\'{\i}}n}, D.J. {Frew},
  M.~{Stupar}, R.~{V{\'a}zquez}, N.J. {Wright}, R.L.M. {Corradi}, R.A.H.
  {Morris}, \mnras \textbf{431}, 279 (2013).
\newblock \doi{10.1093/mnras/stt160}

\bibitem{VPHAS}
J.E. {Drew}, E.~{Gonzalez-Solares}, R.~{Greimel}, M.J. {Irwin},
  A.~{K{\"u}pc{\"u} Yoldas}, J.~{Lewis}, G.~{Barentsen}, J.~{Eisl{\"o}ffel},
  H.J. {Farnhill}, W.E. {Martin}, J.R. {Walsh}, N.A. {Walton}, M.~{Mohr-Smith},
  R.~{Raddi}, S.E. {Sale}, N.J. {Wright}, P.~{Groot}, M.J. {Barlow}, R.L.M.
  {Corradi}, J.J. {Drake}, J.~{Fabregat}, D.J. {Frew}, B.T. {G{\"a}nsicke},
  C.~{Knigge}, A.~{Mampaso}, R.A.H. {Morris}, T.~{Naylor}, Q.A. {Parker},
  S.~{Phillipps}, C.~{Ruhland}, D.~{Steeghs}, Y.C. {Unruh}, J.S. {Vink},
  R.~{Wesson}, A.A. {Zijlstra}, \mnras \textbf{440}, 2036 (2014).
\newblock \doi{10.1093/mnras/stu394}

\bibitem{UWIFE}
J.J. {Lee}, B.C. {Koo}, Y.H. {Lee}, H.G. {Lee}, J.H. {Shinn}, H.J. {Kim},
  Y.~{Kim}, T.S. {Pyo}, D.S. {Moon}, S.C. {Yoon}, M.Y. {Chun}, D.~{Froebrich},
  C.J. {Davis}, W.P. {Varricatt}, J.~{Kyeong}, N.~{Hwang}, B.G. {Park}, M.G.
  {Lee}, H.M. {Lee}, M.~{Ishiguro}, \mnras \textbf{443}, 2650 (2014).
\newblock \doi{10.1093/mnras/stu1146}

\bibitem{mathewson64}
D.S. {Mathewson}, J.R. {Healey}, in \emph{The Galaxy and the Magellanic
  Clouds}, \emph{IAU Symposium}, vol.~20, ed. by F.J. {Kerr} (1964), \emph{IAU
  Symposium}, vol.~20, p. 283

\bibitem{westerlund66}
B.E. {Westerlund}, D.S. {Mathewson}, \mnras \textbf{131}, 371 (1966)

\bibitem{russell90}
S.C. {Russell}, M.A. {Dopita}, \apjs \textbf{74}, 93 (1990).
\newblock \doi{10.1086/191494}

\bibitem{payne08}
J.L. {Payne}, G.L. {White}, M.D. {Filipovi{\'c}}, \mnras \textbf{383}, 1175
  (2008).
\newblock \doi{10.1111/j.1365-2966.2007.12620.x}

\bibitem{chu97}
Y.H. {Chu}, \aj \textbf{113}, 1815 (1997).
\newblock \doi{10.1086/118393}

\bibitem{mathewson80}
D.S. {Mathewson}, M.A. {Dopita}, I.R. {Tuohy}, V.L. {Ford}, \apjl \textbf{242},
  L73 (1980).
\newblock \doi{10.1086/183406}

\bibitem{brantseg14}
T.~{Brantseg}, R.L. {McEntaffer}, L.M. {Bozzetto}, M.~{Filipovic},
  N.~{Grieves}, \apj \textbf{780}, 50 (2014).
\newblock \doi{10.1088/0004-637X/780/1/50}

\bibitem{seward84}
F.D. {Seward}, F.R. {Harnden}, Jr., D.J. {Helfand}, \apjl \textbf{287}, L19
  (1984).
\newblock \doi{10.1086/184388}

\bibitem{LMC_pulsar}
{Fermi LAT Collaboration}, M.~{Ackermann}, A.~{Albert}, L.~{Baldini},
  J.~{Ballet}, G.~{Barbiellini}, C.~{Barbieri}, D.~{Bastieri}, R.~{Bellazzini},
  E.~{Bissaldi}, R.~{Bonino}, E.~{Bottacini}, T.J. {Brandt}, J.~{Bregeon},
  P.~{Bruel}, R.~{Buehler}, G.A. {Caliandro}, R.A. {Cameron}, P.A. {Caraveo},
  C.~{Cecchi}, E.~{Charles}, A.~{Chekhtman}, C.C. {Cheung}, J.~{Chiang},
  G.~{Chiaro}, S.~{Ciprini}, J.~{Cohen-Tanugi}, A.~{Cuoco}, S.~{Cutini},
  F.~{D{`}Ammando}, F.d.P.R. {Desiante}, S.W. {Digel}, L.~{Di Venere}, P.S.
  {Drell}, C.~{Favuzzi}, S.J. {Fegan}, E.C. {Ferrara}, A.~{Franckowiak},
  S.~{Funk}, P.~{Fusco}, F.~{Gargano}, D.~{Gasparrini}, N.~{Giglietto},
  F.~{Giordano}, G.~{Godfrey}, I.A. {Grenier}, M.H. {Grondin}, J.E. {Grove},
  L.~{Guillemot}, S.~{Guiriec}, K.~{Hagiwara}, A.K. {Harding}, E.~{Hays}, J.W.
  {Hewitt}, A.B. {Hill}, D.~{Horan}, T.J. {Johnson}, J.~{Kn{\"o}dlseder},
  M.~{Kuss}, S.~{Larsson}, L.~{Latronico}, M.~{Lemoine-Goumard}, J.~{Li},
  L.~{Li}, F.~{Longo}, F.~{Loparco}, M.N. {Lovellette}, P.~{Lubrano},
  S.~{Maldera}, A.~{Manfreda}, F.~{Marshall}, P.~{Martin}, M.~{Mayer}, M.N.
  {Mazziotta}, P.F. {Michelson}, N.~{Mirabal}, T.~{Mizuno}, M.E. {Monzani},
  A.~{Morselli}, I.V. {Moskalenko}, S.~{Murgia}, G.~{Naletto}, E.~{Nuss},
  T.~{Ohsugi}, M.~{Orienti}, E.~{Orlando}, D.~{Paneque}, M.~{Pesce-Rollins},
  F.~{Piron}, G.~{Pivato}, T.A. {Porter}, S.~{Rain{\`o}}, R.~{Rando},
  M.~{Razzano}, A.~{Reimer}, O.~{Reimer}, T.~{Reposeur}, R.W. {Romani}, P.M.S.
  {Parkinson}, A.~{Schulz}, C.~{Sgr{\`o}}, E.J. {Siskind}, D.A. {Smith},
  F.~{Spada}, G.~{Spandre}, P.~{Spinelli}, D.J. {Suson}, H.~{Takahashi}, J.B.
  {Thayer}, D.J. {Thompson}, L.~{Tibaldo}, D.F. {Torres}, Y.~{Uchiyama},
  G.~{Vianello}, K.S. {Wood}, M.~{Wood}, L.~{Zampieri}, Science \textbf{350},
  801 (2015).
\newblock \doi{10.1126/science.aac7400}

\bibitem{cline80}
T.L. {Cline}, U.D. {Desai}, G.~{Pizzichini}, B.J. {Teegarden}, W.D. {Evans},
  R.W. {Klebesadel}, J.G. {Laros}, K.~{Hurley}, M.~{Niel}, G.~{Vedrenne}, \apjl
  \textbf{237}, L1 (1980).
\newblock \doi{10.1086/183221}

\bibitem{kulkarni03}
S.R. {Kulkarni}, D.L. {Kaplan}, H.L. {Marshall}, D.A. {Frail}, T.~{Murakami},
  D.~{Yonetoku}, \apj \textbf{585}, 948 (2003).
\newblock \doi{10.1086/346110}

\bibitem{guver12}
T.~{G{\"u}ver}, E.~{G{\"o}{\v g}{\"u}{\c s}}, F.~{{\"O}zel}, \mnras
  \textbf{424}, 210 (2012).
\newblock \doi{10.1111/j.1365-2966.2012.21184.x}

\bibitem{rest05}
A.~{Rest}, N.B. {Suntzeff}, K.~{Olsen}, J.L. {Prieto}, R.C. {Smith}, D.L.
  {Welch}, A.~{Becker}, M.~{Bergmann}, A.~{Clocchiatti}, K.~{Cook}, A.~{Garg},
  M.~{Huber}, G.~{Miknaitis}, D.~{Minniti}, S.~{Nikolaev}, C.~{Stubbs}, \nat
  \textbf{438}, 1132 (2005).
\newblock \doi{10.1038/nature04365}

\bibitem{danziger76}
I.J. {Danziger}, M.~{Dennefeld}, \apj \textbf{207}, 394 (1976).
\newblock \doi{10.1086/154507}

\bibitem{vogt11}
F.~{Vogt}, M.A. {Dopita}, \apss \textbf{331}, 521 (2011).
\newblock \doi{10.1007/s10509-010-0479-7}

\bibitem{seok13}
J.Y. {Seok}, B.C. {Koo}, T.~{Onaka}, \apj \textbf{779}, 134 (2013).
\newblock \doi{10.1088/0004-637X/779/2/134}

\bibitem{mathewson72}
D.S. {Mathewson}, J.N. {Clarke}, \apjl \textbf{178}, L105 (1972).
\newblock \doi{10.1086/181095}

\bibitem{seward81}
F.D. {Seward}, M.~{Mitchell}, \apj \textbf{243}, 736 (1981).
\newblock \doi{10.1086/158641}

\bibitem{payne07}
J.L. {Payne}, G.L. {White}, M.D. {Filipovi{\'c}}, T.G. {Pannuti}, \mnras
  \textbf{376}, 1793 (2007).
\newblock \doi{10.1111/j.1365-2966.2007.11561.x}

\bibitem{blair00}
W.P. {Blair}, J.A. {Morse}, J.C. {Raymond}, R.P. {Kirshner}, J.P. {Hughes},
  M.A. {Dopita}, R.S. {Sutherland}, K.S. {Long}, P.F. {Winkler}, \apj
  \textbf{537}, 667 (2000).
\newblock \doi{10.1086/309077}

\bibitem{crawford14}
E.J. {Crawford}, M.D. {Filipovi{\'c}}, R.L. {McEntaffer}, T.~{Brantseg},
  K.~{Heitritter}, Q.~{Roper}, F.~{Haberl}, D.~{Uro{\v s}evi{\'c}}, \aj
  \textbf{148}, 99 (2014).
\newblock \doi{10.1088/0004-6256/148/5/99}

\bibitem{dodorico78}
S.~{Dodorico}, P.~{Benvenuti}, F.~{Sabbadin}, \aap \textbf{63}, 63 (1978)

\bibitem{long90}
K.S. {Long}, W.P. {Blair}, R.P. {Kirshner}, P.F. {Winkler}, \apjs \textbf{72},
  61 (1990).
\newblock \doi{10.1086/191409}

\bibitem{gordon98}
S.M. {Gordon}, R.P. {Kirshner}, K.S. {Long}, W.P. {Blair}, N.~{Duric}, R.C.
  {Smith}, \apjs \textbf{117}, 89 (1998).
\newblock \doi{10.1086/313107}

\bibitem{williams15}
B.F. {Williams}, B.~{Wold}, F.~{Haberl}, K.~{Garofali}, W.P. {Blair}, T.J.
  {Gaetz}, K.D. {Kuntz}, K.S. {Long}, T.G. {Pannuti}, W.~{Pietsch}, P.P.
  {Plucinsky}, P.F. {Winkler}, \apjs \textbf{218}, 9 (2015).
\newblock \doi{10.1088/0067-0049/218/1/9}

\bibitem{dodorico82}
S.~{D'Odorico}, W.M. {Goss}, M.A. {Dopita}, \mnras \textbf{198}, 1059 (1982).
\newblock \doi{10.1093/mnras/198.4.1059}

\bibitem{kumar76}
C.K. {Kumar}, \pasp \textbf{88}, 323 (1976).
\newblock \doi{10.1086/129950}

\bibitem{dodorico80}
S.~{Dodorico}, M.A. {Dopita}, P.~{Benvenuti}, \aaps \textbf{40}, 67 (1980)

\bibitem{blair81}
W.P. {Blair}, R.P. {Kirshner}, R.A. {Chevalier}, \apj \textbf{247}, 879 (1981).
\newblock \doi{10.1086/159098}

\bibitem{braun93}
R.~{Braun}, R.A.M. {Walterbos}, \aaps \textbf{98}, 327 (1993)

\bibitem{magnier95}
E.A. {Magnier}, S.~{Prins}, J.~{van Paradijs}, W.H.G. {Lewin}, R.~{Supper},
  G.~{Hasinger}, W.~{Pietsch}, J.~{Truemper}, \aaps \textbf{114}, 215 (1995)

\bibitem{pietsch05}
W.~{Pietsch}, M.~{Freyberg}, F.~{Haberl}, \aap \textbf{434}, 483 (2005).
\newblock \doi{10.1051/0004-6361:20041990}

\bibitem{sasaki12}
M.~{Sasaki}, W.~{Pietsch}, F.~{Haberl}, D.~{Hatzidimitriou}, H.~{Stiele},
  B.~{Williams}, A.~{Kong}, U.~{Kolb}, \aap \textbf{544}, A144 (2012).
\newblock \doi{10.1051/0004-6361/201219025}

\bibitem{dickel82}
J.R. {Dickel}, S.~{Dodorico}, M.~{Felli}, M.~{Dopita}, \apj \textbf{252}, 582
  (1982).
\newblock \doi{10.1086/159584}

\bibitem{sjouwerman01}
L.O. {Sjouwerman}, J.R. {Dickel},  \textbf{565}, 433 (2001).
\newblock \doi{10.1063/1.1377132}

\bibitem{galvin14}
T.J. {Galvin}, M.D. {Filipovic}, Serbian Astronomical Journal \textbf{189}, 15
  (2014).
\newblock \doi{10.2298/SAJ140505002G}

\bibitem{fesen15_m31}
R.A. {Fesen}, P.A. {H{\"o}flich}, A.J.S. {Hamilton}, \apj \textbf{804}, 140
  (2015).
\newblock \doi{10.1088/0004-637X/804/2/140}

\bibitem{lee15_m81}
M.G. {Lee}, J.~{Sohn}, J.H. {Lee}, S.~{Lim}, I.S. {Jang}, Y.~{Ko}, B.C. {Koo},
  N.~{Hwang}, S.C. {Kim}, B.G. {Park}, \apj \textbf{804}, 63 (2015).
\newblock \doi{10.1088/0004-637X/804/1/63}

\bibitem{payne04}
J.L. {Payne}, M.D. {Filipovi{\'c}}, T.G. {Pannuti}, P.A. {Jones}, N.~{Duric},
  G.L. {White}, S.~{Carpano}, \aap \textbf{425}, 443 (2004).
\newblock \doi{10.1051/0004-6361:20041058}

\bibitem{pannuti02}
T.G. {Pannuti}, N.~{Duric}, C.K. {Lacey}, A.M.N. {Ferguson}, M.A. {Magnor},
  C.~{Mendelowitz}, \apj \textbf{565}, 966 (2002).
\newblock \doi{10.1086/337918}

\bibitem{pannuti07}
T.G. {Pannuti}, E.M. {Schlegel}, C.K. {Lacey}, \aj \textbf{133}, 1361 (2007).
\newblock \doi{10.1086/510718}

\bibitem{kronberg85}
P.P. {Kronberg}, P.~{Biermann}, F.R. {Schwab}, \apj \textbf{291}, 693 (1985).
\newblock \doi{10.1086/163108}

\bibitem{fenech08}
D.M. {Fenech}, T.W.B. {Muxlow}, R.J. {Beswick}, A.~{Pedlar}, M.K. {Argo},
  \mnras \textbf{391}, 1384 (2008).
\newblock \doi{10.1111/j.1365-2966.2008.13986.x}

\bibitem{chevalier01}
R.A. {Chevalier}, C.~{Fransson}, \apjl \textbf{558}, L27 (2001).
\newblock \doi{10.1086/323569}

\bibitem{kuntz10}
K.D. {Kuntz}, S.L. {Snowden}, \apjs \textbf{188}, 46 (2010).
\newblock \doi{10.1088/0067-0049/188/1/46}

\bibitem{danny12}
D.~{Milisavljevic}, R.A. {Fesen}, R.A. {Chevalier}, R.P. {Kirshner},
  P.~{Challis}, M.~{Turatto}, \apj \textbf{751}, 25 (2012).
\newblock \doi{10.1088/0004-637X/751/1/25}

\bibitem{danny08}
D.~{Milisavljevic}, R.A. {Fesen}, \apj \textbf{677}, 306 (2008).
\newblock \doi{10.1086/528929}

\bibitem{badenes10}
C.~{Badenes}, D.~{Maoz}, B.T. {Draine}, \mnras \textbf{407}, 1301 (2010).
\newblock \doi{10.1111/j.1365-2966.2010.17023.x}

\bibitem{dwarkadas05}
V.V. {Dwarkadas}, \apj \textbf{630}, 892 (2005).
\newblock \doi{10.1086/432109}

\bibitem{chomiuk09}
L.~{Chomiuk}, E.M. {Wilcots}, \apj \textbf{703}, 370 (2009).
\newblock \doi{10.1088/0004-637X/703/1/370}

\bibitem{thompson09}
T.A. {Thompson}, E.~{Quataert}, N.~{Murray}, \mnras \textbf{397}, 1410 (2009).
\newblock \doi{10.1111/j.1365-2966.2009.14889.x}

\bibitem{reynolds81}
S.P. {Reynolds}, R.A. {Chevalier}, \apj \textbf{245}, 912 (1981).
\newblock \doi{10.1086/158868}

\bibitem{berezhko04}
E.G. {Berezhko}, H.J. {V{\"o}lk}, \aap \textbf{427}, 525 (2004).
\newblock \doi{10.1051/0004-6361:20041111}

\bibitem{clarke76}
J.N. {Clarke}, \mnras \textbf{174}, 393 (1976)

\bibitem{blair85}
W.P. {Blair}, R.P. {Kirshner}, \apj \textbf{289}, 582 (1985).
\newblock \doi{10.1086/162919}

\bibitem{hughes84}
J.P. {Hughes}, D.J. {Helfand}, S.M. {Kahn}, \apjl \textbf{281}, L25 (1984).
\newblock \doi{10.1086/184277}

\bibitem{pavlovic14}
M.Z. {Pavlovic}, A.~{Dobardzic}, B.~{Vukotic}, D.~{Urosevic}, Serbian
  Astronomical Journal \textbf{189}, 25 (2014).
\newblock \doi{10.2298/SAJ1489025P}

\bibitem{urosevic05}
D.~{Uro{\v s}evi{\'c}}, T.G. {Pannuti}, N.~{Duric}, A.~{Theodorou}, \aap
  \textbf{435}, 437 (2005).
\newblock \doi{10.1051/0004-6361:20042535}

\bibitem{clark76}
D.H. {Clark}, J.L. {Caswell}, \mnras \textbf{174}, 267 (1976)

\bibitem{sklovskii60}
I.S. {Shklovskii}, Soviet Astronomy \textbf{4}, 355 (1960)

\bibitem{case98}
G.L. {Case}, D.~{Bhattacharya}, \apj \textbf{504}, 761 (1998).
\newblock \doi{10.1086/306089}

\bibitem{pavlovic13}
M.Z. {Pavlovi{\'c}}, D.~{Uro{\v s}evi{\'c}}, B.~{Vukoti{\'c}}, B.~{Arbutina},
  {\"U}.D. {G{\"o}ker}, \apjs \textbf{204}, 4 (2013).
\newblock \doi{10.1088/0067-0049/204/1/4}

\bibitem{berkuijsen86}
E.M. {Berkhuijsen}, \aap \textbf{166}, 257 (1986)

\bibitem{rest08}
A.~{Rest}, T.~{Matheson}, S.~{Blondin}, M.~{Bergmann}, D.L. {Welch}, N.B.
  {Suntzeff}, R.C. {Smith}, K.~{Olsen}, J.L. {Prieto}, A.~{Garg}, P.~{Challis},
  C.~{Stubbs}, M.~{Hicken}, M.~{Modjaz}, W.M. {Wood-Vasey}, A.~{Zenteno},
  G.~{Damke}, A.~{Newman}, M.~{Huber}, K.H. {Cook}, S.~{Nikolaev}, A.C.
  {Becker}, A.~{Miceli}, R.~{Covarrubias}, L.~{Morelli}, G.~{Pignata},
  A.~{Clocchiatti}, D.~{Minniti}, R.J. {Foley}, \apj \textbf{680}, 1137 (2008).
\newblock \doi{10.1086/587158}

\bibitem{hughes95}
J.P. {Hughes}, I.~{Hayashi}, D.~{Helfand}, U.~{Hwang}, M.~{Itoh},
  R.~{Kirshner}, K.~{Koyama}, T.~{Markert}, H.~{Tsunemi}, J.~{Woo}, \apjl
  \textbf{444}, L81 (1995).
\newblock \doi{10.1086/187865}

\bibitem{li11}
W.~{Li}, J.~{Leaman}, R.~{Chornock}, A.V. {Filippenko}, D.~{Poznanski},
  M.~{Ganeshalingam}, X.~{Wang}, M.~{Modjaz}, S.~{Jha}, R.J. {Foley},
  N.~{Smith}, \mnras \textbf{412}, 1441 (2011).
\newblock \doi{10.1111/j.1365-2966.2011.18160.x}

\bibitem{winkler85}
P.F. {Winkler}, R.P. {Kirshner}, \apj \textbf{299}, 981 (1985).
\newblock \doi{10.1086/163764}

\bibitem{morse95}
J.A. {Morse}, P.F. {Winkler}, R.P. {Kirshner}, \aj \textbf{109}, 2104 (1995).
\newblock \doi{10.1086/117436}

\bibitem{athena}
X.~{Barcons}, K.~{Nandra}, D.~{Barret}, J.W. {den Herder}, A.C. {Fabian},
  L.~{Piro}, M.G. {Watson}, {the Athena Team}, Journal of Physics Conference
  Series \textbf{610}(1), 012008 (2015).
\newblock \doi{10.1088/1742-6596/610/1/012008}

\bibitem{badenes09}
C.~{Badenes}, J.~{Harris}, D.~{Zaritsky}, J.L. {Prieto}, \apj \textbf{700}, 727
  (2009).
\newblock \doi{10.1088/0004-637X/700/1/727}

\bibitem{jennings12}
Z.G. {Jennings}, B.F. {Williams}, J.W. {Murphy}, J.J. {Dalcanton}, K.M.
  {Gilbert}, A.E. {Dolphin}, M.~{Fouesneau}, D.R. {Weisz}, \apj \textbf{761},
  26 (2012).
\newblock \doi{10.1088/0004-637X/761/1/26}

\bibitem{jennings14}
Z.G. {Jennings}, B.F. {Williams}, J.W. {Murphy}, J.J. {Dalcanton}, K.M.
  {Gilbert}, A.E. {Dolphin}, D.R. {Weisz}, M.~{Fouesneau}, \apj \textbf{795},
  170 (2014).
\newblock \doi{10.1088/0004-637X/795/2/170}

\end{thebibliography}
\bibliographystyle{spphys}

\printindex
\end{document}